\newif\ifconfver
\confvertrue        

\newif\ifonecoltab

\newif\ifplainver  

\ifplainver
    \confverfalse                         
\fi
\ifconfver
     \documentclass[10pt,journal]{IEEEtran}
\else
    \ifplainver
        \documentclass[11pt]{article}
        \usepackage{fullpage}
    \else
        \documentclass[12pt,draftcls,onecolumn]{IEEEtran}
    \fi
\fi

\usepackage{calc,amsfonts,amssymb,amsmath,bm,url,color,theorem,graphicx,cite,epstopdf,nicefrac,bbold}
\usepackage{psfrag,float,subfig}
\usepackage[algoruled,linesnumbered]{algorithm2e}
\usepackage{multirow}
\usepackage[normalem]{ulem}
\usepackage{stmaryrd}
\usepackage{xcolor}
\usepackage{psfrag,framed,mdframed}
\usepackage{comment}
\usepackage{mdframed}
\usepackage{threeparttable}
\usepackage{bibunits}

\definecolor{orange}{RGB}{255,107,0}
\def\blue{\color{blue}}

\newcommand{\W}{\boldsymbol{W}}

\newcommand{\X}{\boldsymbol{X}}
\newcommand{\C}{\boldsymbol{C}}

\renewcommand{\H}{\boldsymbol{H}}

\newcommand{\V}{\boldsymbol{V}}

\newcommand{\x}{\boldsymbol{x}}

\newcommand{\z}{\boldsymbol{z}}
\renewcommand{\c}{\boldsymbol{c}}

\newcommand{\w}{\boldsymbol{w}}

\newcommand{\T}{{\!\top\!}}
\DeclareMathOperator*{\argmin}{\arg\min}

\DeclareMathOperator*{\minimize}{\textrm{minimize}}

\newtheorem{definition}{Definition}

\usepackage{xcolor}
\usepackage{psfrag,framed}
\usepackage{lipsum}
\usepackage{chapterbib}
\PassOptionsToPackage{normalem}{ulem}
\usepackage{ulem}
\definecolor{shadecolor}{RGB}{220,220,220}

\usepackage{import,mathtools,listings,wrapfig,hhline,ulem,afterpage}
\definecolor{green}{rgb}{0,0.5686,0.1529 }
\definecolor{purple}{rgb}{0.72,0,0.69}

\newcommand{\norm}[1]{\left\lVert#1\right\rVert}
\newcommand{\set}[1]{\left\{#1\right\}}
\newcommand\numberthis{\addtocounter{equation}{1}\tag{\theequation}} 
\newcommand{\abs}[1]{\left\lvert#1\right\rvert}

\renewcommand{\exp}[1]{\textnormal{exp}(#1)}
\floatname{algorithm}{Algorithm}

\DeclareMathOperator*{\argmax}{\arg\max}

\newtheorem{Fact}{Fact}
\newtheorem{Lemma}{Lemma}
\newtheorem{Prop}{Proposition}
\newtheorem{Theorem}{Theorem}

\theorembodyfont{\rmfamily}

\newtheorem{Assumption}{Assumption}

\usepackage{lipsum}
\usepackage{amsfonts}
\usepackage{graphicx}
\usepackage{epstopdf}
\ifpdf
  \DeclareGraphicsExtensions{.eps,.pdf,.png,.jpg}
\else
  \DeclareGraphicsExtensions{.eps}
\fi

\def\blue{\color{blue}}

\usepackage{xcolor,etoolbox}

\begin{document}

\newcommand{\papertitle}{
Memory-Efficient Convex Optimization for Self-Dictionary Separable Nonnegative Matrix Factorization: A Frank-Wolfe Approach
}

\newcommand{\paperabstract}{%
Nonnegative matrix factorization (NMF) often relies on the {\it separability} condition for tractable algorithm design. Separability-based NMF is mainly handled by two types of approaches, namely, greedy pursuit and convex programming.
A notable convex NMF formulation is the so-called { self-dictionary multiple measurement vectors} (SD-MMV), which can work without knowing the matrix rank {\it a priori}, and is arguably more resilient to error propagation relative to greedy pursuit. However, convex SD-MMV renders a large memory cost that scales {\it quadratically} with the problem size. This memory challenge has been around for a decade, and a major obstacle for applying convex SD-MMV to big data analytics. This work proposes a memory-efficient algorithm for convex SD-MMV. 
Our algorithm capitalizes on the special update rules of a classic algorithm from the 1950s, namely, the Frank-Wolfe (FW) algorithm.
It is shown that, under reasonable conditions, the FW algorithm solves the noisy SD-MMV problem with a memory cost that grows {\it linearly} with the amount of data. To handle noisier scenarios, a smoothed group sparsity regularizer is proposed to improve robustness while maintaining the low memory footprint with guarantees. 
The proposed approach presents the first linear memory complexity algorithmic framework for convex SD-MMV based NMF. The method is tested over a couple of unsupervised learning tasks, i.e., text mining and community detection, to showcase its effectiveness and memory efficiency.
}


\ifplainver

    \date{\today}

    \title{\papertitle}

    \author{
    Tri Nguyen, Xiao Fu, and Ruiyuan Wu\\
    School of Electrical Engineering and Computer Science\\
    Oregon State University\\
    Email: (nguyetr9, xiao.fu)@oregonstate.edu
    }

	\date{}

    \maketitle

\else
    \title{\papertitle}

    \ifconfver \else {\linespread{1.1} \rm \fi

\author{Tri Nguyen, Xiao Fu, and Ruiyuan Wu
	
	\thanks{

    	This work is supported in part by the Army Research Office (ARO) under Project ARO W911NF19-1-0407.
		
		T. Nguyen and X. Fu are with the School of Electrical Engineering and Computer Science, Oregon State University, Corvallis, OR 97331, United States. Email:  (nguyetr9, xiao.fu)@oregonstate.edu
		
		R. Wu is with the Department of Electronic Engineering, The Chinese University of Hong Kong, Shatin, N.T., Hong Kong SAR. Emai: rywu@cuhk.edu.hk

	}
}

\maketitle

\ifconfver \else
\begin{center} \vspace*{-2\baselineskip}
\end{center}
\fi

\begin{abstract}
	\paperabstract
\end{abstract}

\begin{IEEEkeywords}\vspace{-0.0cm}%
Unsupervised multimodal analysis, sample complexity, identifiability
\end{IEEEkeywords}

    \ifconfver \else \IEEEpeerreviewmaketitle} \fi

 \fi

\ifconfver \else
    \ifplainver \else
        \newpage
\fi \fi

\section{Introduction}
Nonnegative matrix factorization (NMF) aims at factoring a data matrix $\X \in\mathbb{R}^{M\times N}$ into a product of two latent nonnegative factor matrices, i.e., 
\begin{equation}
    \label{eq:nmf}
  \X\approx \W\H, \quad  \W\in\mathbb{R}_+^{M\times K},~\H\in\mathbb{R}_+^{K\times N},
\end{equation}
where $K\leq \min\{M,N\}$.
The NMF technique has been a workhorse for dimensionality reduction, representation learning, and blind source separation.
It finds a plethora of applications in machine learning and signal processing; see, e.g., \cite{fu2018nonnegative,gillis2020nonnegative}. 
In particular, NMF plays an essential role in many statistical model learning problems, e.g., topic modeling \cite{arora2012practical,huang2016anchor,fu2018anchor}, community detection \cite{huang2019detecting,mao2017mixed,panov2018consistent,ibrahim2020mixed}, crowdsourced data labeling \cite{ibrahim2021crowdsourcing,ibrahim2019crowdsourcing}, and joint probability estimation \cite{ibrahim2021recovering}.

One major challenge of NMF lies in computation.
It was shown in \cite{vavasis2009complexity} that NMF is an NP-hard problem in the worst case, even without noise. In the last two decades, many approximate algorithms were proposed to tackle the NMF problem; see \cite{fu2020computing,fu2018nonnegative,gillis2020nonnegative}. Notably, a line of work exploiting physically reasonable assumptions to come up with {\it polynomial-time} NMF algorithms has drawn considerable attention. To be specific, the so-called {\it separable} NMF approaches leverage the {\it separability} condition
to design efficient and tractable algorithms. More importantly, separable NMF algorithms are often provably robust to noise \cite{fu2015robust,gillis2014robust,fu2014self,arora2012practical,gillis2014fast}.

The separability condition is a special sparsity-related condition. Interestingly, it is nonetheless well-justified in many applications. 
For example, in topic modeling, the separability condition holds if every topic has an ``anchor word'' that does not appear in other topics \cite{arora2012learning}. 
In community detection, separability translates to the existence of a set of ``pure nodes'' whose membership is only associated with a specific community \cite{mao2017mixed}. 
The ``pure pixel assumption'' in hyperspectral imaging is also identical to separability, which means that there exist pixels that only capture one material spectral signature \cite{ma2014asignal,MC01}. In crowdsourced data labeling, separability is equivalent to the existence of expert annotators who are specialized for one class \cite{ibrahim2021crowdsourcing}.

Two major categories of algorithms exist for separable NMF. The first category is greedy algorithms. The representative algorithm is the {\it successive projection algorithm} (SPA), which was first proposed in the hyperspectral imaging community \cite{MC01}, and was re-discovered a number of times from different perspectives; see \cite{ren2003automatic,arora2012practical,fu2014self,chan2011simplex,nascimento2005vertex,fu2015blind,fu2013greedy}. Many of these greedy algorithms have a Gram-Schmidt structure, and thus are very scalable. However, they also share the same challenge of Gram-Schmidt, i.e., error propagation. The second category formulates the separable NMF problem as all-at-once convex optimization criteria (see, e.g., \cite{esser2012convex,recht2012factoring,fu2015robust,Elhamifar2012,gillis2018afast,mizutani2014ellipsoidal,arora2012learning,gillis2014robust}), which are arguably more robust to noise and less prone to error propagation.

Among the all-at-once convex formulations of separable NMF, the {\it self-dictionary multiple measurement vectors} (SD-MMV) based framework in \cite{esser2012convex,recht2012factoring,fu2015robust,Elhamifar2012,gillis2018afast,gillis2014robust} has a series of appealing features, e.g., 
identifiability of the latent factors without knowing the model rank, computational tractability, and noise robustness.
Nonetheless, this line of work has serious challenges in terms of memory.
These algorithms induce a variable that has a size of $N\times N$, which is not possible to instantiate if $N$ reaches the level of $N=100,000$---which leads to a $74.5$GB matrix if the double precision is used.
Consequently, these algorithms are often used together with a data selection pre-processing stage to reduce problem size (see, e.g., \cite{gillis2018afast}). This may again create an error propagation issue---if the data selection stage missed some important data samples (e.g., those related to anchor words or pure nodes in topic modeling and community detection, respectively), the algorithms are bound to fail.

\smallskip

\noindent
{\bf Contributions.}
In this work, we revisit convex optimization-based SD-MMV for separable NMF. 
Our goal is to offer an NMF identifiability-guaranteed algorithm that is also provably memory-efficient. 
Our contributions are as follows:

\noindent
$\bullet$ {\bf A Memory-Efficient Algorithmic Framework.} We first show that applying with the standard FW algorithm onto a special SD-MMV formulation for separable NMF admits identifiability of the ground-truth latent factors, if the noise is absent. More importantly, the memory cost of this algorithm is $O(KN)$ other than $O(N^2)$, where $K\ll N$ often holds.
Based on this simple observation, we further show that similar conclusions can be drawn even if noise is present---under more conditions.

\noindent
$\bullet$ {\bf Regularization-Based Performance Enhancement.}
To circumvent stringent conditions and to improve noise robustness,
we propose a smoothed group-sparsity regularization that is reminiscent of the mixed norm regularization often used in SD-MMV formulations (see \cite{esser2012convex,fu2015robust,fu2015power,Elhamifar2012,ammanouil2014blind}). We show that the optimal solution of such regularized formulation corresponds to the desired ground-truth NMF factors in the noisy case---i.e., identifiability of the NMF holds. We further show that this regularization can better safeguard the memory consumption within the range of $O(KN)$ compared to the unregularized version, if the FW algorithm is initialized reasonably. 
To our best knowledge, the proposed FW algorithmic framework is the first convex SD-MMV method whose memory cost scales {\it linearly} in $N$, while existing methods such as those in \cite{gillis2014robust,fu2015robust,recht2012factoring,esser2012convex,gillis2018afast} all need $O(N^2)$ memory.

\noindent
$\bullet$ {\bf Synthetic and Real data Evaluation.}
We test the proposed approach on various synthetic and real datasets. In particular, we evaluate our algorithm using a couple of text corpora, i.e., NIST Topic Detection and Tracking (TDT2) {\cite{fiscus1999nist}} and the Reuters-21578 data\footnote{ http://www.daviddlewis.com/resources/testcollections/reuters21578} on {\it topic mining} tasks.
We also use social network data from \cite{ley2002dblp}, \cite{sinha2015overview} to evaluate the performance on mixed membership {\it community detection} tasks. The proposed approach is benchmarked by competitive state-of-the-art baselines for solving separable NMF.

\smallskip

\noindent
{\bf Notation.} We follow the established conventions in signal processing. In particular,
$\bm{x} \in \mathbb{R}^{N}$ denotes a real-valued $N$-dimensional vector; $x_n$ and $[\x]_n$ both denote the $n$th element of $\x$;
$\bm{X} \in \mathbb{R}^{N \times M}$ denotes a real-valued matrix;
$\text{rank}(\cdot)$ denotes matrix rank;
superscript $^\T$ is used for transpose; $\|\cdot\|_{p}$ where $p \geq 1$ denotes the vector  $\ell_p$ norm; $\|\cdot\|_{\rm F}$ denotes the Frobenius norm; $\bm{X}(n, :)$ and $\x_\ell$ denote  the $n$th row and  $\ell$th column of $\bm{X}$, respectively; 
$x_{n, \ell}$, $[\X]_{n,\ell}$ and $\X(n,\ell)$ all denote the $(n,\ell)$th element of $\X$;  $\bm{X}(-n, :)$ and $\bm{X}(:, -\ell)$ denote the submatrices constructing from $\X$ by removing the $n$th row and  $\ell$th column, respectively; $|\mathcal{K}|$ and $ \mathcal{K}^{c}$ denote the cardinality and complement of the set ${\cal K}$, respectively; $\norm{\bm{X}}_{{\rm row}-0}$ counts the number of nonzero rows of $\X$; $ \norm{\bm{X}}_{\infty, 1}= \sum^{M}_{i=1} \norm{\bm{X}(i, :)}_{\infty}$ is the mixed $\ell_\infty/\ell_1$ norm;
${\rm supp}(\bm{x})$ denotes a set of all indices of non-zero elements of vector $ \bm{x}$, i.e., $ {\rm supp}(\bm{x}) \coloneqq \set{i \mid [\bm{x}]_i \neq 0}$; ${\rm conv}\{\bm{x}_1, \bm{x}_2, \ldots , \bm{x}_k\}$ denotes convex hull of set $\set{\bm{x}_1, \bm{x}_2, \ldots , \bm{x}_k}$;
$\bm{1}$ and $\bm{0}$ denote an all-one matrix/vector and an all-zero matrix/vector, respectively, with proper sizes;
$\bm{e}_i$ denotes the $i$th unit vector; 
$\bm{I}_N$ denotes an identity matrix with a size of $N\times N$; 
both notations $\bm{x} \geq 0$ and $\bm{X} \geq 0$ mean that the nonnegativity is applied element-wise; $i \in [N]$ means $i\in\{1,\ldots,N\}$; $\sigma_{\max}(\bm{X})$ and $\sigma_{\min}(\bm{X})$ denote the largest and smallest singular value of $\bm{X}$, respectively;
$\lambda_{\max}(\bm{X})$ denotes the largest eigenvalue of $\bm{X}$;
$\mathbb{N}$ denotes the set of natural numbers.

\section{Problem Statement}

Consider a noisy NMF model, i.e.,
\begin{equation}\label{eq:nmf_noise}
    \bm X =\W\H + \V,~\W\geq\bm 0,~\H\geq \bm 0,
\end{equation}
where $\W\in\mathbb{R}^{M\times K}$ and $\H\in\mathbb{R}^{K\times N}$ are nonnegative latent factors as defined before,
and
$\V$ is a noise term. We further assume that 
\begin{equation}\label{eq:sumtoone}
\bm 1^\T\bm H=\bm 1^\T,    
\end{equation}
i.e., the columns of $\H$ reside in the probability simplex. This assumption naturally holds in many applications, e.g., topic modeling, community detection, and hyperspectral unmixing \cite{arora2012learning,fu2016robust,huang2019detecting,ma2014asignal}. When $\H$ does not have sum-to-one columns, this assumption can be ``enforced'' through column normalization of $\X$, under the condition that $\W$ is nonnegative, and there is no noise; see \cite{fu2018nonnegative,gillis2014fast}. 
We should mention that although our interest lies in NMF, the proposed method can also be applied to the so-called {\it simplex-structured matrix factorization}, where $\W$ is not required to be nonnegative; see, e.g., \cite{fu2016robust,ibrahim2020mixed,panov2018consistent,fu2015blind}.

Estimating the ground-truth $\W$ and $\H$ from $\X$ is in general an NP-hard problem \cite{vavasis2009complexity}. However, if the so-called {\it separability} condition holds, the separable NMF algorithms are often tractable, even under noise.

\subsection{Separable NMF}
The separable NMF problem uses the following premise:
\begin{Assumption}
[Separability]\label{as:sep}
There exists a set $${\cal K}=\{\ell_1,\ldots,\ell_K\}$$ such that $\bm H(:,{\cal K})=\bm I_K$. Equivalently, we have $\X(:,{\cal K})=\W$ under \eqref{eq:sumtoone}, if noise is absent.
\end{Assumption}
The condition was first seen in \cite{donoho2003does} in the NMF literature. The remote sensing community discovered it even earlier \cite{MC01}, where the same condition is called the ``pure pixel condition'' \cite{ma2014asignal}. The condition has many names in applications, e.g., the ``anchor word assumption'' in topic modeling \cite{huang2016anchor,fu2018anchor,arora2012practical}, the ``pure node condition'' in community detection \cite{panov2018consistent,mao2017mixed,huang2019detecting}, and the ``specialized annotator condition'' in crowdsourcing \cite{ibrahim2019crowdsourcing}; also see its usage in speech processing \cite{fu2015blind}, image analysis \cite{CANMS} and wireless communications \cite{fu2015power}. 

Under separability, the NMF task boils down to finding ${\cal K}$, since $\W\approx \X(:,{\cal K})$ if the noise level is not high. Then, $\H$ can be estimated by (constrained) least squares if ${\rm rank}(\W)=K$.

\subsection{Convex Separable NMF}

\begin{figure}[t]
    \centering
    \def\svgwidth{1\linewidth}
    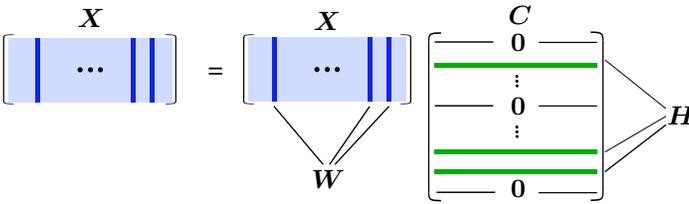
    \caption{Visualization of the idea of the row-sparsity-based SD-MMV in \cite{esser2012convex,fu2014self}.} 
    \label{fig:sdmmv-demonstration}
\end{figure}

A way to look at the ${\cal K}$-finding problem is to cast it as a sparse atom selection problem. To be specific, when noise is absent, consider the following row-sparsity minimization formulation:
\begin{subequations}\label{eq:self}
\begin{align}
    \minimize_{\C\in\mathbb{R}^{N\times N}}&~\|\C\|_{{\rm row}-0}\\
    {\rm subject~to}&~\X=\X\C,~\C\geq \bm 0,~\bm 1^\T\bm C=\bm 1^\T,
\end{align}
\end{subequations}
where $\|\C\|_{{\rm row}-0}$ counts the nonzero rows of $\C$. For example, if  ${\cal K}=\{1,\ldots,K\}$, then an optimal solution of \eqref{eq:self} is 
\begin{equation}
\C^\star = \begin{bmatrix} \H\\
\bm 0\end{bmatrix}    
\end{equation}
under mild conditions (e.g., ${\rm rank}(\W)=K$). 
More formally, we have: 
\begin{equation}\label{eq:selfopt}
   \C^\star({\cal K},:)=\H,~\C^\star({\cal K}^c,:)=\bm 0;
\end{equation}
see a proof for the general case in \cite{fu2014self} and an illustration in Fig.~\ref{fig:sdmmv-demonstration}. Hence, ${\cal K}$ can be identified by inspecting the nonzero rows of $\C^\star$. The formulation in \eqref{eq:self} is reminiscent of the multiple measurement vectors (MMV) problem from compressive sensing \cite{tropp2006algorithms,Chen2006}, but using the data itself as the dictionary---which is the reason why \eqref{eq:self} is called {\it self-dictionary multiple measurement vectors} (SD-MMV) \cite{fu2014self}. We should mention that SD-MMV can be regarded as a way of picking up the vertices of the convex hull of $\{\x_1,\ldots,\x_N\}$, which is a popular perspective that many separable NMF algorithms take; see, e.g., \cite{nascimento2005vertex,chan2011simplex,arora2012practical,winter1999nfindr}. Unlike the classic vertex picking methods that often need the knowledge of $K$, SD-MMV can work without knowing the number of vertices.

The formulation in \eqref{eq:self} is not easy to tackle, due to the combinatorial nature of $\|\C\|_{{\rm row}-0}$. One popular way is to use greedy pursuit, which selects the basis that represents $X$ using its convex combinations from the self-dictionary $\X$ one by one. This leads to the {\it successive projection algorithm} (SPA) \cite{fu2014self,gillis2014fast,MC01,chan2011simplex}. The greedy pursuit methods are often effective, efficient, and robust to a certain extent. However, they also share the same challenge---the error could be accumulated and propagated through the greedy steps.

Another line of work employs the convex relaxation idea and use a convex surrogate for  $\|\C\|_{{\rm row}-0}$---see \cite{fu2015robust,gillis2013robustness,esser2012convex,recht2012factoring,fu2015power,Elhamifar2012,gillis2018afast,ammanouil2014blind} for different options. For example, the work in \cite{esser2012convex,fu2015robust} uses
\begin{equation}
    \label{eq:ideal_solution}
      \|\C\|_{\infty,1} =\sum_{n=1}^N\|\C(n,:)\|_\infty
\end{equation}
as their convex surrogate. When the noise is present, the work in \cite{esser2012convex,fu2015robust} also advocated the following formulation
\begin{subequations}\label{eq:self-fitting}
\begin{align}
    \minimize_{\C\in\mathbb{R}^{N\times N}}&~\frac{1}{2}\|\X-\X\C\|^2_{\rm F}+\lambda\|\C\|_{\infty,1}\\
    {\rm subject~to}&~\C\geq \bm 0,~\bm 1^\T\bm C=\bm 1^\T,
\end{align}
\end{subequations}
where $\lambda\geq 0$ is a regularization parameter. 
The formulation is convex and thus is {\it conceptually} easy to handle.
After using any convex optimization algorithm to obtain an optimal solution $\widehat{\C}$,
the ${\cal K}$ set can be identified via inspecting $\|\widehat{\C}(n,:)\|_\infty$ for all $n$ and picking the $K$ indices that are associated with $K$ largest row $\ell_\infty$-norms. This method was shown to have identifiability of ${\cal K}$ under noise \cite{fu2015robust}---also see similar results for close relatives of \eqref{eq:self-fitting} in \cite{gillis2013robustness,gillis2014robust,recht2012factoring,gillis2018afast}.

In terms of noise robustness, the convex optimization-based SD-MMV methods are arguably more preferable over greedy methods, since they identify all elements of ${\cal K}$ simultaneously, instead of in a one by one manner that is prone to error propagation.

\subsection{The Memory Challenge} 
Using convex programming to handle separable NMF is appealing, since many off-the-shelf algorithms are readily available. The convergence properties of convex optimization algorithms are also well understood. 
However, general-purpose convex optimization algorithms, e.g., the interior-point methods and gradient descent, may encounter serious challenges when handling Problem~\eqref{eq:self-fitting}.
The reason is that the $\C$ variable induces $O(N^2)$ memory if a general-purpose solver is used. 
This often makes running such algorithms costly and slow, even when $N$ is only moderate. 

Many attempts were made towards accelerating convex SD-MMV.
The early work in \cite{recht2012factoring} uses a fast projection to expedite the algorithm for a variant of \eqref{eq:self-fitting}. The recent work in \cite{gillis2018afast} employed the fast gradient paradigm for accelerated convergence. However, the $O(N^2)$ memory issue could not be circumvented in both approaches.
In this work, our objective is a convex optimization algorithm that share similar identification properties of those in \cite{fu2015robust} but has provable memory-efficiency---i.e., only $O(NK)$ memory is needed for instantiating the optimization variable $\C$.

\section{A Frank-Wolfe Approach}
\label{sec:unreg}

\subsection{Preliminaries on Frank-Wolfe Algorithm} \label{sec:prelim}
Our development is based on the idea of the Frank-Wolfe (FW) algorithm that was developed in the 1950s \cite{frank1956algorithm}. The FW algorithm deals with problems of the following form:
\begin{subequations}
\begin{align}
    \minimize_{\bm \theta \in\mathbb{R}^d}&\quad f(\bm \theta)\\
    {\rm subject~to}&\quad \bm \theta\in {\cal C},
\end{align}
\end{subequations}
where ${\cal C}$ is compact and convex and $f(\cdot):\mathbb{R}^d\rightarrow \mathbb{R}$ is differentiable and convex. 

The FW algorithm uses the following updates:
\begin{subequations}\label{eq:FW}
\begin{align}
       \bm u &\leftarrow \arg\min_{\bm u\in {\cal C}}~\nabla f({\boldsymbol \theta^{t}})^\T \bm u, \label{eq:fw_u}\\
        \bm \theta^{t+1} &\leftarrow (1-\alpha^t)\bm \theta^t + \alpha^t \bm u.
\end{align}
\end{subequations}
Here,
$\alpha^t$ is a pre-defined sequence, i.e.,
\begin{equation}\label{eq:alphaschedule}
     \alpha^t = \frac{2}{t+2},~t=0, 1, 2, \ldots.
\end{equation}
The above two steps constitute a standard procedure of the FW algorithm \cite{jaggi2013revisiting}, \cite{frank1956algorithm}. The idea of FW is intuitive: In each iteration, FW linearizes the cost function and solve it locally over the compact constraint set.
For convex problems, the plain-vanilla FW algorithm converges to an $\varepsilon$-optimal solution using $O(1/\varepsilon)$ iterations (which is also known as a sublinear convergence rate). Recent works showed that FW (and its variants) converges even faster (with a linear rate) under some more conditions; see, e.g., \cite{jaggi2013revisiting,freund2016new}.

When dealing with large-scale optimization, especially memory-wise costly optimization problems, FW may help circumvent memory explosion due to its special update rule---which has already facilitated economical nuclear norm optimization paradigms that are important in recommender systems \cite{jaggi2013revisiting}.
FW also features simple updates if the constraint set is
\[ {\cal C}=\{\bm \theta\in\mathbb{R}^N ~|~ \bm 1^\T\bm \theta=1,\bm \theta\geq \bm 0 \}. \]
In this case, $\bm u$ in \eqref{eq:FW} is always a unit vector. This is because \eqref{eq:fw_u} is a linear program over the probability simplex ${\cal C}$, and the minimum is always attained at a vertex of ${\cal C}$ \cite{boyd2004convex}. The vertices of ${\cal C}$ are $\{\bm e_n\}_{n=1}^N$.
Hence, the solution of \eqref{eq:fw_u} is $\bm e_j$ where
\begin{equation}\label{eq:fw_u_sln}
      j = \arg\min_{n\in[N]} [\nabla f(\bm \theta^t)]_{n} .  
\end{equation}
In our algorithm design, we will take advantage of this fact to reduce the memory cost of solving SD-MMV.

Note that one needs not always start the FW algorithm from $t=0$. In \cite{freund2016new}, a {\it warm-start based Frank-Wolfe} (WS-FW) algorithm was proposed for accelerated convergence. There, one can use $\bm \theta^{t_{\rm init}}=\bm \theta^{{\rm init}}$ with $t_{\rm init}\geq 1$ and start the FW algorithm from the $t_{\rm init}$th iteration using the corresponding $\alpha^{t_{\rm init}}$ as defined in \eqref{eq:alphaschedule}. By carefully choosing $t_{\rm init}$ according to the quality of $\bm \theta^{{\rm init}}$ and the problem structure (e.g., the curvature of the cost function), WS-FW is shown to converge to the global optimum more efficiently.

\color{black}

\subsection{Warm-up: The Noiseless SD-MMV}
\label{subsec:warm-up}
Our algorithm design starts with the simple formulation as follows:
\begin{subequations}\label{eq:unreg}
\begin{align}
    \minimize_{\C}&~\dfrac{1}{2}\|\X-\X\C\|_{\rm F}^2\\
    {\rm subject~to}&~\bm 1^\T\bm C=\bm 1^\T,~\bm C\geq \bm 0.
\end{align}
\end{subequations}
Note that in general, \eqref{eq:unreg} does not admit identifiability of ${\cal K}$---the optimal solution of \eqref{eq:unreg} needs not to have the form in \eqref{eq:selfopt}. A simple counter example is $\C=\bm I_N$---which gives zero objective value, but is not the desired solution. That is, one cannot infer ${\cal K}$ from the nonzero rows of $\bm C=\bm I_N$.

\subsection{Noiseless Case and Simple Self-Dictionary Fitting}
\label{subsec:noiseless}
The first observation is that although the criterion in \eqref{eq:unreg} does not admit identifiability of ${\cal K}$, 
the FW algorithm guarantees finding a $\C$ as in \eqref{eq:selfopt}, thereby pining down ${\cal K}$.

To see this, note that the updates of $\bm{c}_\ell$'s do not affect each other.
Consider the following update rule for $\ell=1,\ldots,N$:
\begin{subequations}\label{eq:FW_ours}
\begin{align}
       j &\leftarrow \arg\min_{n\in[N]} [\X^\T ( \bm{X} \c_\ell^t - \x_\ell)]_n, \label{eq:updating_rule1}\\
       \bm c_\ell^{t+1} &\leftarrow (1-\alpha^t)\c_\ell^t + \alpha^t \bm e_j, \label{eq:updating_rule}
\end{align}
\end{subequations}
Since the formulation is convex, the above updates are guaranteed to solve the problem in \eqref{eq:unreg}.
Before we examine what solution the updates will lead to, let us discuss its memory complexity, i.e., the reason why such updates are potentially memory-economical.

To see how much memory that the algorithm needs beyond storing $\bm{X}$ (which is inevitable), let us analyze the steps in \eqref{eq:FW_ours}. 
First, evaluating gradient in \eqref{eq:updating_rule1} is not memory-costing. One can always evaluate $\bm{c}_\ell^{t} - \bm{x}_\ell$, followed by multiplying with $\bm{X}$, followed by another left multiplying $\bm{X}^\T$. These three operations produce vectors as their results and hence require $O(N)$ memory complexity in total. Note that one can evaluate the gradient w.r.t. $\C$ column by column, and no previously evaluated columns need to be stored. Hence, the total memory cost is $O(N)$. Second, the remaining memory requirement lies in storing the iterates $\bm{c}_\ell^{t+1}$ for $\ell=1,\ldots,N$. Consider an ideal case where the $j$ found in \eqref{eq:updating_rule} satisfies $j\in{\cal K}$ for all $\ell$ and all $t$. Then, if the initialization $\bm{C}^{0} = \bm{0}$, the updating rule in \eqref{eq:updating_rule} always maintains ${\rm supp}(\bm{c}_\ell^t)\subseteq {\cal K}$ for all $\ell=1,\ldots,N$. This leads to an $O(KN)$ memory for storing $\bm{C}^t$. 

Based on our discussion, the key to attain the above described memory efficiency is to ensure that $j \in \mathcal{K}$ for all $\ell$ and all $t$ in \eqref{eq:updating_rule}. We show that this is indeed the case under some conditions:
\begin{Theorem}[Noiseless Case, Memory Efficiency] \label{theorem:noiseless}
    Suppose that Assumption \ref{as:sep} holds, that ${\rm rank}(\W)=K$, that no repeated unit vectors exist in $\H$\footnote{In applications where there are many identical/similar columns in $\H$ (e.g., hyperspectral unmixing), the assumption can be ``enforced'' by clustering-based pre-processing \cite{esser2012convex,gillis2018afast}.},
and that the noise is absent (i.e., $\V=\bm 0$). 
Furthermore, define $\bm{q}_\ell^t = \bm{W}^\T \bm{W}(\bm{H}\bm{c}_\ell^t - \bm{h}_\ell)$.
Assume that the following holds:
\begin{align}\label{eq:nodupq}
 q_{i,\ell}^t - \min_{j} q_{j,\ell}^t \neq 0,~\forall i \neq \argmin_{j} q_{j,\ell}^t,
\end{align}
before $\bm c_\ell^t$ reaches an optimal solution.
Then, the FW algorithm in \eqref{eq:FW} with initialization $ \bm{C}^0 = \bm{0}$ always outputs the desired solution in \eqref{eq:selfopt} using $O(KN)$ memory beyond storing $\X$.
\end{Theorem}
The proof is relegated to Appendix~\ref{app:noiseless}.
Note that Theorem~\ref{theorem:noiseless} requires that Eq.~\eqref{eq:nodupq} holds---i.e., no repeated minimum-value elements of $\bm q_\ell$ exist for any $\bm c_\ell^t$. This is not hard to meet under some mild conditions:
\begin{Fact}\label{fact:prob}
Assume that the columns of $\bm W$ follow any joint continuous distribution. Then, we have
\[  {\sf Pr}( q_{i,\ell} =q_{j,\ell}) =0,~\forall i,j\in[K],\ell\in[N]. \]
\end{Fact}
\begin{IEEEproof}
Under the assumption, suppose that there exists any $b$ such that we have ${\sf Pr}(\w_i^\T\bm z=b,\w_j^\T\z=b)>0$ for any vector $\bm z$. That is, there is a positive probability that $\bm q_\ell$ could have two identical elements.
However, this cannot happen. Indeed, note that $y_i=\w_i^\T\bm z$ and $y_j=\w_j^\T\z$ are also continuous random variables. Hence, ${\sf Pr}(y_i=b,y_j=b)=0$ for any $b$. This means that 
${\sf Pr}(\w_i^\T\bm z=\w_j^\T\z)=0$.
Letting $\bm z= \W(\H\c_\ell^t-\bm h_\ell)$ and applying the above result complete the proof.
\end{IEEEproof}
Although Theorem~\ref{theorem:noiseless} is concerned with the ideal noiseless case, which is hardly practical,
the observation in Theorem~\ref{theorem:noiseless} opens another door for self-dictionary convex optimization-based NMF.
That is, it shows that using FW to solve the self-dictionary problem may successfully circumvent the memory explosion issue.

\subsection{The Noisy Case}
\label{sub:noise_unreg}
A natural question lies in if the same procedure works in the presence of noise. The answer is affirmative. To understand this, we first notice that in the noiseless case, we have
\[
    \H\bm c_\ell = \bm h_\ell,~{\rm supp}(\bm c_\ell) \subseteq {\cal K}~\forall \ell {\in [N]}\Longrightarrow \bm c_\ell = \bm c_\ell^\star, { \forall \ell \in [N]}.
\]
where $\bm{c}_\ell^{\star}$ is the $\ell$th column  in $\bm{C}^{\star}$ defined in \eqref{eq:selfopt}.
This further leads to
\[
    \norm{\bm{C}(n, :)}_{\infty} =1, \forall n\in \mathcal{K},\quad \norm{\bm{C}(n, :)}_{\infty} = 0, \forall n \in \mathcal{K}^{c},
\] 
which is the reason why one can easily infer ${\cal K}$ from the solution $\C=\C^\star$.
When noise is present, FW may not be able to find the exact $\bm{C}^{\star}$. 
However, an approximate $\C\approx \C^\star$ often suffices to identify ${\cal K}$. 
In the following, Lemma~\ref{lem:alpha} shows that instead of finding $\c_\ell$ such that $\bm{H}\bm{c}_\ell=\bm{h}_\ell$, seeking a $\c_\ell$ such that $\bm{H}\bm{c}_\ell\approx \bm{h}_\ell$ using FW does not hurt recovering $\mathcal{K}$, under reasonable conditions:
\begin{Lemma}
    \label{lem:alpha}
    Assume that the separability condition holds, and that $\; {\rm rank}(\W)=K$, and that no repeated unit vectors appear in $\H$.
    Suppose that a feasible solution $\C$ satisfies
    ${\rm supp}(\c_\ell)\subseteq{\cal K}$ and $\norm{\bm{H}\bm{c}_\ell - \bm{h}_\ell}_2 \leq \omega$ for all $\ell\in[N]$.
    Then, we have
    \begin{align*}
        & \norm{\bm{C}(n, :)}_{\infty} \geq  1 - \omega \sqrt{K}/2, \; n \in \mathcal{K},\\
        &\norm{\bm{C}(n, :)}_{\infty} = 0, \;  n \notin \mathcal{K};
    \end{align*}
   i.e., ${\cal K}$ can still be identified by inspecting $\|\C(n,:)\|_\infty$, if $\omega<2/\sqrt{K}$.
\end{Lemma}
\begin{IEEEproof}
    Let $\ell \in \mathcal{K}$ , we have
    \begin{align*}
        \omega \geq  \norm{\H\c_\ell - \bm h_\ell}_2 
               &= \norm{ \sum_{i \in \mathcal{K} } c_{i,\ell} \bm{e}_i  - \bm{e}_k }_2 \text{for some  $k \in \mathcal{K}$ }  \\
          & \geq \dfrac{1}{\sqrt{K}} \norm{ \sum_{i \in \mathcal{K} } c_{i,\ell}  \bm{e}_i  - \bm{e}_k }_1 \\
         & = \dfrac{1}{\sqrt{K}}\norm{(c_{k,\ell} - 1)\bm{e}_k + \sum_{i \neq k} c_{i,\ell}  \bm{e}_i }_1 \\
          &= \dfrac{1}{\sqrt{K}} \left(  (1 - c_{k,\ell} ) + \sum_{i \neq k} c_{i,\ell}  \right) \\
&          = \dfrac{2}{\sqrt{K}}(1 - c_{k,\ell} ) \quad \text{(since  $\bm{1}^\T \bm{c}_\ell = 1$)}.
    \end{align*}
    \[
        \Longrightarrow \norm{\bm{C}(k, :)}_{\infty} \geq c_{k,\ell}  \geq 1 - \omega \dfrac{\sqrt{K}}{2} \; \text{for $k \in \mathcal{K}$.}
    \] 

    On the other hands, since ${\rm supp}(\bm{c}_\ell) \subseteq \mathcal{K}$ and $\bm{C}$ is initialized at $\bm{0}$, we have
    \[
         \norm{\bm{C}(n, :)}_{\infty} = 0 \quad \quad \text{for $n \in \mathcal{K}^{c}$}.
    \] 
    This completes the proof.
\end{IEEEproof}

Next, we show in the following theorem that even in the noisy case, applying FW onto \eqref{eq:unreg} produces a solution that is a reasonable approximation of $\bm{C}^{\star}$.
To proceed, let us define the following quantities:
\begin{definition}
    \label{def:set1}
    Define the following terms:
    \begin{align*}
        \gamma &\coloneqq \max_{1\leq k \leq K} \norm{\bm{w}_k}_2, ~
d(\bm{H}) \coloneqq \max_{n \in \mathcal{K}, \ell \in \mathcal{K}^{c}} h_{n, \ell}, \\
        \delta  &\coloneqq \max_{1 \leq i\leq N } \norm{\bm{v}_i}_2.
    \end{align*} 
\end{definition}
In particular, a small $d(\H)$ means that all $\bm h_\ell$'s for $\ell\in{\cal K}^c$ are sufficiently different from the unit vectors. This is a desirable case, since small perturbation would not confuse such $\bm h_\ell$'s and the unit vectors.

\begin{Theorem}[Noisy Case, Memory Efficiency]
\label{theorem:noisy_unreg}
Suppose that Assumption \ref{as:sep} holds, that there is no repeated unit vector in $\H$, and that ${\rm rank}(\W)=K$.
    Furthermore, let $\widetilde{\bm{q}}^t = \bm{W}^\T \bm{W} (\bm{H}\bm{c}_\ell^t - \bm{h}_\ell) / \norm{\bm{H}\bm{c}_\ell - \bm{h}_\ell}_2$.
    During FW's updates, assume a positive gap between the smallest and second smallest elements of $\widetilde{\bm{q}}^t$ exists, i.e., 
    \begin{align}\label{eq:wiredcond}
       \min_{\ell \in [N], t }~ \left(\widetilde{q}_{i,\ell}^t - \min_{j} \widetilde{q}_{j,\ell}^t\right) \geq \nu,~\forall i \neq \argmin_{j} \widetilde{q}_{j,\ell}^t,     
    \end{align}
    before the FW algorithm terminates. Also assume that
    {
    \begin{equation}
    \label{eq:noisy_unreg_cond}
    \delta \leq \sqrt{\gamma^2 + \dfrac{\nu \eta (1-d(\bm{H}))}{4\sigma_{\max}(\bm{W})}} - \gamma,
    \end{equation} 
    }
    for some $\eta > 0$. {Then, if one terminates FW when $\|\bm x_\ell-\X\c_\ell^t\|_2\leq\eta + 2 \delta$,} the algorithm produces a solution $\widehat{\bm{C}}$ that satisfies $$\norm{\bm{H}\widehat{\bm{c}}_\ell - \bm{h}_\ell}_2 \leq ( \eta + 4 \delta )/\sigma_{\min}(\bm{W}). $$
    In addition, during the process, ${\rm supp}(\bm c_\ell^t)\subseteq {\cal K}$ always holds for all $t$ and $\ell$, and thus only $O(KN)$ memory is taken for instantiating $\bm{C}^{t}$ for all $t$.
\end{Theorem}
The proof can be found in Appendix~\ref{appedix:proof_theorem_noisy_unreg}.
Theorem~\ref{theorem:noisy_unreg} shows that under certain conditions, the FW algorithm for solving \eqref{eq:unreg} indeed gives a reasonable solution $\widehat{\bm C}$ using only $O(NK)$ memory. Combining with Lemma~\ref{lem:alpha}, one can see that ${\cal K}$ can be correctly selected if the noise level is not high.
Both Theorems \ref{theorem:noiseless} and \ref{theorem:noisy_unreg} reveal an unconventional identifiability result. Note that the formulation in \eqref{eq:unreg} {\it per se} does not have identifiability of $\bm C^\star$. That is, the optimal solutions of \eqref{eq:unreg} do not necessarily reveal ${\cal K}$---as mentioned, $\bm I_N$ is also an optimal solution. However, when one uses a particular algorithm (i.e., FW) to solve \eqref{eq:unreg}, the produced solution sequence converges to a ${\cal K}$-revealing $\widehat{\bm C}$, even if noise is present.

On the other hand, the identifiability and memory efficiency come with caveats. First, the condition in \eqref{eq:wiredcond} is hard to check or guarantee.
As seen in Fact~\ref{fact:prob}, $\nu>0$ does exist under mild conditions---but the quantity of this parameter may change from instance to instance and is hard to acquire. 
Using the formulation in \eqref{eq:unreg} is also not the most ``natural'', since we know that the desired $\C^\star$ should be very row-sparse---why not using this information for performance enhancement.
Can we get rid of the condition in \eqref{eq:wiredcond} and effectively use the prior knowledge on $\C^\star$? The answer is affirmative, with a re-design of the objective function.
In the next section, we will discuss these aspects.

\section{Performance Enhancement via Regularization}
Under the formulation in \eqref{eq:unreg}, the gap specified in \eqref{eq:wiredcond} was essential for the FW to pick $j\in{\cal K}$ in every step.
In this section, our idea is to use a regularization term to help the FW algorithm to achieve the same goal while not relying on the condition in \eqref{eq:wiredcond}. In addition, since the desired $\C^\star$ in \eqref{eq:selfopt} has a row-sparse structure, it is natural to add a regularization to exploit this prior information. 

Using row-sparsity-promoting regularization terms is a common practice for self-dictionary convex optimization-based NMF; see, e.g., \cite{fu2015robust,gillis2014robust,recht2012factoring,esser2012convex,fu2015power,gillis2018afast,Elhamifar2012,ammanouil2014blind}. 
In particular, \cite{fu2015power,esser2012convex,Elhamifar2012,ammanouil2014blind,fu2015robust} all used the popular convex mixed norms such as $\|\C\|_{\infty,1}$ or $\|\C\|_{q,1}$ (where $q\geq 1$) for row-sparsity encouraging---which are reminiscent of compressive sensing \cite{tropp2006algorithms,Chen2006}.
However, such convex norms may not be the best to combine with our FW framework---since FW arguably works the best with differentiable objectives due to the use of gradient. There are subgradient versions of FW for nonsmooth cost functions (see, e.g.,  \cite{thekumparampil2020projection}), but the algorithmic structure is less succinct. 

We still hope to use a regularization term like $\|\C\|_{\infty,1}$, which was shown to have nice identifiability properties in SD-MMV \cite{esser2012convex,fu2015robust}.
To make this mixed-norm based nonsmooth row-sparsity regularization ``FW-friendly'', we use the following lemma:
\begin{Lemma}{\cite{nesterov2005smooth}}
    \label{lemma:max_smooth}
    For $\mu > 0 $ and $\bm{x} \in \mathbb{R}_+^{N}$, define 
    $ \varphi_{\mu}(\bm{x}) = \mu \log \left(  \nicefrac{1}{N} \sum^{N}_{i=1} \exp{x_i/\mu} \right) $.
    Then function $\varphi_{\mu}(\bm{x})$ is a smooth approximation of $\norm{\bm{x}}_{\infty}$, i.e.,
    \begin{equation*}
    \begin{aligned}
                       &\lim_{\mu \rightarrow 0} \varphi_{\mu}(\bm{x}) = \norm{\bm{x}}_{\infty}  \\ 
                       &\norm{\bm{x}}_{\infty} - \mu \log(N) \leq \varphi_{\mu}(\bm{x}) \leq \norm{\bm{x}}_{\infty}.
    \end{aligned}
    \end{equation*}
\end{Lemma}
The above smoothing technique is from \cite{nesterov2005smooth}. A proof is presented in Appendix~\ref{proof:lemmax_max_smooth} in the supplementary material for completeness.
Building upon Lemma~\ref{lemma:max_smooth}, a smooth approximation of $\norm{\bm{C}}_{\infty, 1}$ is readily obtained as 
$$  \varPhi_{\mu}(\bm{C}) = \sum^{N}_{n = 1} \varphi_\mu(\bm{C}(n, :)) \approx \norm{\bm{C}}_{\infty, 1}.$$ 
Using $\varPhi_{\mu}(\bm{C})$, our working formulation is as follows:
\begin{mdframed}
\begin{subequations}\label{eq:self-fw-fitting}
\begin{align}
    \minimize_{\C\in\mathbb{R}^{N\times N}}&\quad\frac{1}{2}\|\X-\X\C\|^2_{\rm F}+\lambda \varPhi_{\mu}(\C)\\
    {\rm subject~to}&\quad\C\geq \bm 0,~\bm 1^\T\bm C=\bm 1^\T.
\end{align}
\end{subequations}
\end{mdframed}
The formulation can be understood as a smoothed version of those in \cite{fu2015robust,fu2015power,esser2012convex,Elhamifar2012}. Note that the problem in \eqref{eq:self-fw-fitting} is still convex, but easier to handle by gradient-based algorithms relative to the nonsmooth version.

\subsection{Identifiability of ${\cal K}$}
Our first step is to understand the optimal solution of \eqref{eq:self-fw-fitting}---i.e., if one optimally solves Problem~\eqref{eq:self-fw-fitting}, is the solution still ${\cal K}$-revealing as in the nonsmooth version?
To this end, we will use the following definition \cite{fu2015robust,gillis2014robust}:
\begin{definition}
    The quantity $\kappa(\bm{W})$ is defined as follows:
    \begin{equation}
        \label{def:alpha}
        \kappa(\bm{W}):= \min_{\substack{k \in [K], 
        \\ \bm{1}^\T\boldsymbol \theta = 1, \boldsymbol \theta \geq 0}} \norm{\bm{w}_k - \bm{W}(:,{-k}) \boldsymbol \theta}_2.
    \end{equation}
\end{definition}
The term $\kappa(\W)$ in a way reflects the ``conditioning'' of $\W$. If $\kappa(\W)$ is large, it means that every $\w_k$ is far away from the convex hull spanned by the remaining columns of $\W$, which implies that ${\rm conv}\{\w_1,\ldots,\w_K\}$ is well-stretched over all directions. This is a desired case, since such convex hulls are more resilient to small perturbations when estimating the vertices (i.e., $\bm w_k$ for $k=1,\ldots,K$).

With this definition, we show that the optimal solution of \eqref{eq:self-fw-fitting} can reveal ${\cal K}$ under some conditions:
\begin{Theorem}[Identifiability]
    \label{theorem:identifibility}
Assume that Assumption \ref{as:sep} holds, that there is no repeated unit vector in $\H$, and that ${\rm rank}(\W)=K$. Also assume that $ \norm{\bm{v}_i}^2 \leq (\rho / N) \norm{\bm{V}}_{\rm F}^2$ for some $\rho$ and  $\norm{\bm{v}_i}_2 \leq \delta$.
Then, any optimal solution $\bm{C}_{\rm opt}$ of Problem \eqref{eq:self-fw-fitting} satisfies: 
\begin{subequations}
\label{eq:reg_iden}
\begin{align} 
    \|\bm{C}_{\rm opt}(n,:)\|_\infty &> 1- \beta, ~ \forall n \in {\cal K}, \label{eq:pure}\\
    \|\C_{\rm opt}(n,:)\|_\infty &\leq 2\rho\dfrac{N-K}{\lambda N} \norm{\bm{V}}_{\rm F}^2 \label{eq:non-pure}\\
    &+ \mu N \log(N) + \beta K ,  ~~\forall n \in {\cal K}^c \nonumber
\end{align}
\end{subequations}
where $
\beta = \nicefrac{ \sqrt{4\rho (1-K/N) \norm{\bm{V}}_{\rm F}^2 + 2\lambda K} + 2\delta }
        { \kappa(\bm{W})(1-d(\bm{H}))}.$ 
\end{Theorem}
The proof is relegated to Appendix~\ref{app:noisyident}.

Theorem~\ref{theorem:identifibility} reveals the ``interplay'' between the noise level and the hyperparameters $\lambda,\mu$. In other words, it states that given a certain noise level, there may exist a pair of $\lambda,\mu$ such that ${\cal K}$ will be identified using the proposed FW algorithm. For example, when the noise level is not high, a natural choice is to concentrate more on the fitting error rather than the regularization term. This is reflected in conditions \eqref{eq:pure} and \eqref{eq:non-pure}.
Since $\norm{\bm{V}}_{\rm F}^2$ is small, $\delta$ would be also small. Then, with a small $\lambda$ to suppress the term $2\lambda K$ in the expression of $\beta$, a small $\beta$ can be expected. Similarly, with a small $\mu$ to suppress the term $\mu N\log N$ in \eqref{eq:non-pure}, the right hand side (RHS) of \eqref{eq:non-pure} would be close to $0$. In addition, when the noise level is relatively high, one would want to increase $\lambda$ to counter the effect of increasing $\norm{\V}_{\rm F}^2$ in \eqref{eq:non-pure}, but not to increase it to an overly large extent (in order to keep $\beta$ close to $0$, due to the presence of $2\lambda K$ in the expression of $\beta$).

Theorem~\ref{theorem:identifibility} asserts that with the proposed regularization, finding an optimal solution of \eqref{eq:self-fw-fitting} is useful for identifying ${\cal K}$. Such optimal solutions can be attained by {\it any} convex optimization algorithm. 
Nonetheless, theorem~\ref{theorem:identifibility} only safeguards the final solution of our formulation in \eqref{eq:self-fw-fitting}, 
which is a similar result as in \cite{fu2015robust} for the nonsmooth regularization version \eqref{eq:self-fitting}.
However, Theorem \ref{theorem:identifibility} does not speak for the memory efficiency.

\subsection{Memory Efficiency of Frank-Wolfe for \eqref{eq:self-fw-fitting}}
In this subsection, we show that running FW to optimize the new objective in \eqref{eq:self-fw-fitting} costs only ${O}(KN)$ memory under some conditions---which are arguably milder compared to the conditions for the unregularized case as shown in Theorem~\ref{theorem:noisy_unreg}.
In the regularized case, the FW updates become the following:
\begin{equation}
        \label{eq:sub_update}
    \begin{aligned}
        j &= \argmin_{n\in[N]} \; [ \bm{g}_\ell]_n, \\ 
        \bm{c}_\ell^{t+1} &= (1- \alpha)\bm{c}_\ell^{t} + \alpha \bm e_j,
    \end{aligned}
 \end{equation}
where $\bm{g}_\ell$ is the $\ell$th column of $\nabla \widehat{f}(\bm{C}^{t})$ in which $\widehat{f}(\bm{C}) \coloneqq \dfrac{1}{2} \norm{\bm{X} - \bm{X}\bm{C}}_{\rm F}^2 + \lambda \varPhi_{\mu}(\bm{C})$.
Similar as before, the key to establish memory efficiency is to show that $j$ picked in \eqref{eq:sub_update} always belongs to ${\cal K}$.
\begin{algorithm}[t]
    \caption{\texttt{MERIT}}
    \label{alg:fw_reg}
    \SetAlgoLined
    \KwIn{$\bm{X} \in \mathbb{R}^{M \times N}, \lambda, t_{\rm init}, \bm{C}^{\rm init} \text{(if Warm Start})$}

    \eIf {`\texttt{Warm~Start}'=\texttt{ True}} {
        $\bm{C} = \bm{C}^{\rm init}$\;
    } {
        $\bm{C} = \bm{0}$\;
        $t_{\rm init} = 0$\;
    }

\For{$t \leftarrow t_{\rm init}, t_{\rm init} +1, t_{\rm init} +2, \ldots $}{

    $\alpha \leftarrow 2 / (t+2)$\; \label{code:start}
    $\bm{r} \leftarrow \bm{0}$\;
    \For{$n \leftarrow 1, \ldots , N$} {
        $r_n \leftarrow \sum^{N}_{i=1} \exp{\bm{C}(n, i)/ \mu}$\;
    }
    \For {$\ell \leftarrow 1, \ldots , N$} {
    $\bm{p}_\ell \leftarrow \bm{X}^\T (\bm{X}\bm{c}_\ell - \bm{x}_\ell)$\;
    $\bm{y}_\ell \leftarrow \exp{\bm{c}_\ell/\mu}\; {\cdot/}\; \bm{r}$  \tcp*{`$\cdot/$' denotes an element-wise division between 2 vectors}
    $\bm{g}_\ell \leftarrow \bm{p}_\ell + \lambda \bm{y}_\ell$\;
    $\bm{u}_\ell \leftarrow \bm{e}_i$ such that $i = \argmin_{j} \; [\bm{g}_\ell]_j$\; \label{code:end}

    $\bm{c}_\ell \leftarrow (1 - \alpha) \bm{c}_\ell + \alpha \bm{u}_\ell $\; \label{code:update}
    
    }
    
}
    \KwOut{$\bm{C}$.}
\end{algorithm}
The next theorem shows that the regularization helps achieve this goal:
\begin{Theorem}[Regularized Case, Memory Efficiency]
    \label{theorem:reg_mem}
Consider the regularized case and the FW algorithm in \eqref{eq:sub_update}.
Assume that Assumption \ref{as:sep} holds and that ${\rm rank}(\W)=K$.
Also assume that $(i)$ in iteration $t$, $\bm{C}^{t}$ satisfies ${\rm supp}(\bm{c}^{t}_\ell) \subseteq \mathcal{K}$ for all $\ell\in[N]$ and $(ii)$ there exists at least an $n^\star \in \mathcal{K}$ such that $\C^t(n^\star,:)$ is not a constant row vector.
If the noise bound satisfies
    \begin{align}
        \label{eq:noise_bound_reg}
         &\delta \leq  
         \sqrt{\gamma^2 + \upsilon
        } -\gamma, 
    \end{align} 
where 
\begin{align*}\upsilon=\dfrac{1}{4}
    (\nicefrac{\lambda}{N} &- \lambda \exp{\nicefrac{-\psi}{ \mu}} \\
&- {(d'(\bm{H})^2+2d'(\bm{H})+5)\lambda_{\max}(\bm{W}^\T \bm{W})/2}) 
\end{align*} in which we have
\begin{align*}
        \psi &\coloneqq  \min_{\substack{1 \leq i,j \leq N, \\ n \in \mathcal{K},  c^t_{n,i} \neq c^t_{n,j}}} \abs{c^t_{n,i} - c^t_{n, j}},  \\
         d'(\bm{H}) &= \sqrt{\max(2(d(\bm{H}) - \nicefrac{1}{2})^2 +\nicefrac{1}{2}, 2(\nicefrac{1}{2} - \nicefrac{1}{K})^2+ \nicefrac{1}{2})}, 
\end{align*} 
then, to attain $\c_\ell^{t+1}$ from $\c_\ell^t$, FW will update the $j_\ell$th element of $\c_\ell^t$ such that $j_\ell\in{\cal K}$ for all $\ell\in[N]$.
\end{Theorem}
The proof can be found in Appendix~\ref{sub:proof_reg_mem}. 
The noise bound in Theorem~\ref{theorem:reg_mem} is arguably more favorable relative to the unregularized case in Theorem~\ref{theorem:noisy_unreg}. The reason is that $\lambda$ can be tuned to compensate noise. This is also intuitive since a larger $\lambda$ means that one has lower confidence in the data quality due to higher noise.

The key condition in Theorem~\ref{theorem:reg_mem} is the existence of $\bm{C}^{t}(n^{\star}, :)$ that is not a constant---which is reflected in $\psi$. 
At first glance, this is hard to guarantee since it is a characterization of $\C^t$.
Nonetheless, we show that if one can properly {\it initialize} the FW algorithm with an initial solution $\C^{\rm init}$ that satisfies a certain regularity condition, the non-constant condition is automatically satisfied---due to the ``predictable'' update rule of FW:
\begin{Prop}[Initialization Condition]
    \label{proposition:row_is_not_constant}
    Let $\bm{C}^{\rm init}$ be a feasible initial solution. 
    Define $$D_{ij}^n(\bm{C}) = \frac{t_{\rm init} (t_{\rm init}+1)}{2}\abs{c_{n,i} - c_{n,j}}$$ for some $t_{\rm init} \in \mathbb{N}, t_{\rm init}  \geq 1$.
    Suppose that for some $ n^{\star} \in \mathcal{K}$, there exists a pair of $i^\star,j^\star$ such that {the following regularity condition is satisfied}:
     \begin{equation}
         \label{eq:init_condition}
         D_{i^\star j^\star}^{n^{\star}}(\bm{C}^{\rm init}) \notin \mathbb{N}.
    \end{equation} 
    Running WS-FW with $ \C^t=\C^{\rm init}$ starting with $t=t_{init}$ for $T$ iterations, the produced solution sequence $\set{\bm{C}}^{t}$ for $t \geq t_{\rm init}+1$ satisfies $\bm{C}^{t}(n^{\star}, :)$ is not a constant row vector and 
    \[
        \min_{\substack{{ n \in \mathcal{K},} i,j \\ c^t_{n,i} \neq c^t_{n,j}}}\abs{c^t_{n,i} - c^t_{n, j}} \geq \dfrac{2 \xi}{T(T+1)}
    \]
    where 
$
        \xi \coloneqq  \min_{
        \substack{n \in \mathcal{K},i,j, 
            z \in \mathbb{N} \\
            D_{i,j}^n(\bm{C}^{\rm init}) \notin \mathbb{N}
        }
    } \abs{D_{i,j}^{n}(\bm{C}^{\rm init}) - z}.
$ 
\end{Prop}
The proof is in Appendix~\ref{app:propinit}. Theorem~\ref{theorem:reg_mem} and Proposition~\ref{proposition:row_is_not_constant} together mean that for a given problem instance and under a certain noise level, there exist proper parameters $\lambda,
\mu$ that guarantee the recovery of ${\cal K}$ using $O(KN)$ memory to instantiate $\C^t$.

Simply speaking, Proposition~\ref{proposition:row_is_not_constant} asserts that, if there exists a row in $\C^{\rm init}(n^\star,:)$ where $n^{\star}\in{\cal K}$ and this row has two elements whose difference is not a natural number, then Theorem~\ref{theorem:reg_mem} holds with in finite iterations. 
Under Theorem~\ref{theorem:reg_mem} and Proposition~\ref{proposition:row_is_not_constant}, it is natural to run the WS-FW with a $t=t_{\rm init}>0$ as in \cite{freund2016new}. Many lightweight algorithms (e.g., the greedy methods) can be employed to provide the initialization.

The condition in Proposition~\ref{proposition:row_is_not_constant} is fairly mild, since it boils down to the existence of two distinctive elements in any row of the initial $\C$.
The Proposition also suggests that using some existing algorithms to initialize the proposed FW algorithm may be appropriate, since one needs at least one $n^\star \in {\rm supp}(\bm C^{\rm init})\cap {\cal K}$ such that the specified conditions are met.
Any greedy algorithm, e.g., \cite{gillis2014fast,fu2014self}, could help offer this $n^\star$ using a couple of iterations.
Although checking the initialization condition is easy,
we should also mention that it may not be necessary to really check it in practice. The reason is that the regularity condition in Proposition~\ref{proposition:row_is_not_constant} is only sufficient---which means that in practice one often needs not to enforce $\bm C^{\rm init}$ to satisfy it. In fact, the FW algorithm works well and maintains a low memory footprint even using $\bm C^{\rm init}=\bm 0$.

\smallskip

To summarize, we present the {\it \uline{me}mory-efficient F\uline{r}ank-Wolfe based nonnegative matrix factor\uline{i}za\uline{t}ion} (\texttt{MERIT}) algorithm in Algorithm~\ref{alg:fw_reg}. An implementation of \texttt{MERIT} can be downloaded from the authors' website\footnote{https://github.com/XiaoFuLab/Frank-Wolfe-based-method-for-SD-MMV}.

\section{Numerical Results}
In this section, we use synthetic and real data experiments to showcase the effectiveness of the proposed FW-based approach.

\subsection{Synthetic Data Simulations}
\label{subsec:synthetic}

We create synthetic data matrices with different $M,N$ and $K$,
under the noisy signal model $\bm{X} = \bm{W}\bm{H} + \bm{V}$.
The matrix $\bm{W}$ is drawn from the uniform distribution $\mathcal{U}(0, 1)$, the first $K$ columns of  $\bm{H}$ are assigned to be an identity matrix (which means that ${\cal K}=\{1,\ldots,K\}$), the $N-K$ remaining columns of $\bm H$ are generated so that every column resides in the probability simplex (see details later).
After adding zero-mean $\sigma^2$-variance Gaussian noise $\bm V$ to $\bm W\bm H$, the columns of the data matrix are then permuted randomly to obtain the final $\X$---this means that ${\cal K}$ for each random trial is different. The signal-to-noise ratio (SNR) used in this section is defined as SNR$=10\log_{10}( \sum^{N}_{\ell=1} \norm{\bm{W}\bm{h}_\ell}_2^2 ) / ( MN\sigma^2 )$dB. 

\noindent
{\bf Baselines and Metric.} We use the \texttt{SPA} algorithm \cite{gillis2014fast,MC01,fu2014self} that is the prominent greedy algorithm for ${\cal K}$-identification in separable NMF as our baseline.
We also employ the \texttt{FastGradient} algorithm \cite{gillis2018afast} that is designed to solve a convex self-dictionary formulation of SD-MMV. The algorithm uses accelerated gradient for fast convergence, and is considered state-of-the-art. 

We apply the algorithms to the problem instances and select ${\cal K}$ from their outputs as follows.
For \texttt{SPA},
we use the first $K$ indices output by the greedy steps to serve as $\mathcal{K}$.
For \texttt{FastGradient}, we use the authors' implementation and its default methods to pick up $\mathcal{K}$. 
Following Theorem \ref{theorem:identifibility}, we select indices of $K$ rows that have $K$ largest $\norm{\bm{C}(n, :)}_{\infty}$ values as an estimation of ${\cal K}$.

To select the hyperparameter $\lambda$ of \texttt{MERIT}, we use an idea similar to that in 
\cite{gillis2018afast}, which is a heuristic that selects $\lambda$ to balance the data fitting residue and the regularization term. In our case, the suggestion is to set $\lambda = \norm{\bm{X} - \bm{X}\bm{C}^{0}}_{\rm F}/\varPhi_{\mu}(\bm{C}^0)$ (or simply $\lambda = \norm{\bm{X} - \bm{X}\bm{C}^0}_{\rm F}/K$ when $K$ is known), where 
$\bm{C}^0$ is an initial solution that can be constructed by some fast separable NMF algorithms, e.g., \texttt{SPA}. 
The parameter $\mu$ is set to be $10^{-5}$ throughout this section unless otherwise specified, as it is fairly inconsequential.
The hyperparameter of \texttt{FastGradient} is chosen by its default heuristic.

We use a number of metrics to evaluate the performance. In particular, we primarily use the {\it success rate} of ${\cal K}$ identification, which is defined as
$$  \text{success rate} = {\sf Pr}( \mathcal{K} = \widehat{\mathcal{K}} ) .$$
In our simulations, the success rate is estimated using $50$ random trials. 
We also adopt two complementary metrics from \cite{gillis2018afast},
namely, the {\it mean-removed spectral angle} ({MRSA})
and the {\it relative approximation error} (RAE).
In a nutshell, the {MRSA} measures how well $\W$ is estimated via $\widehat{\W}=\X(:,\widehat{\cal K})$ and the RAE measures how well the estimated $\widehat{\W}$ (together with an estimated $\widehat{\H}$) can reconstruct the data $\X$; see details in \cite{gillis2018afast}.
Following \cite{gillis2018afast}, the {MRSA} values are normalized to an interval of $[0, 100]$. In addition, the {RAE} values are in between 0 and 1.
Lower {MRSA}s and higher {RAE}s correspond to better performance of separable NMF.

\noindent
{\bf Results.}
We first evaluate the algorithms under a setting from \cite[Sec. 4.1]{gillis2018afast}, where
$\H(:,{\cal K}^c)$'s columns are middle points between the extreme points of the unit simplex.
This way, noise could easily confuse the $\x_\ell$'s associated with the middle points with the true extreme points of ${\rm conv}\{\x_1,\ldots,\x_N\}$---thereby presenting a challenging case for separable NMF algorithms.

\begin{figure}[t]
    \centering
    \includegraphics[width=0.4\textwidth]{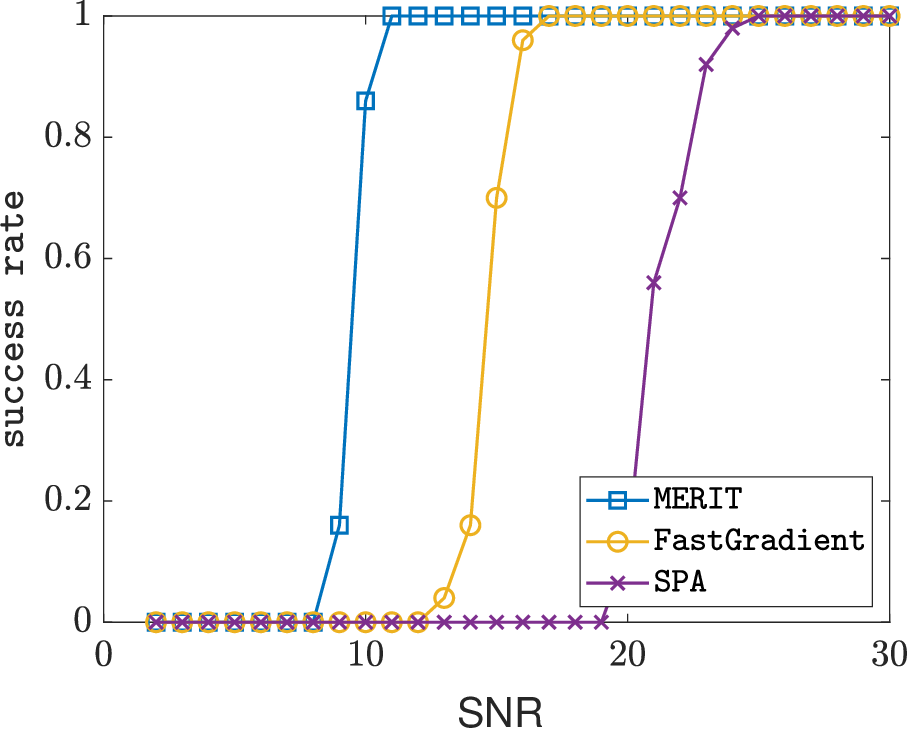}
    \caption{Success rates of the algorithms under various SNRs; $(M,N,K)=(50,55,10)$.}\label{fig:ratenew} 
\end{figure}

\begin{figure}[t]
    \centering
    \subfloat[MRSA]{
        \includegraphics[width=0.23\textwidth]{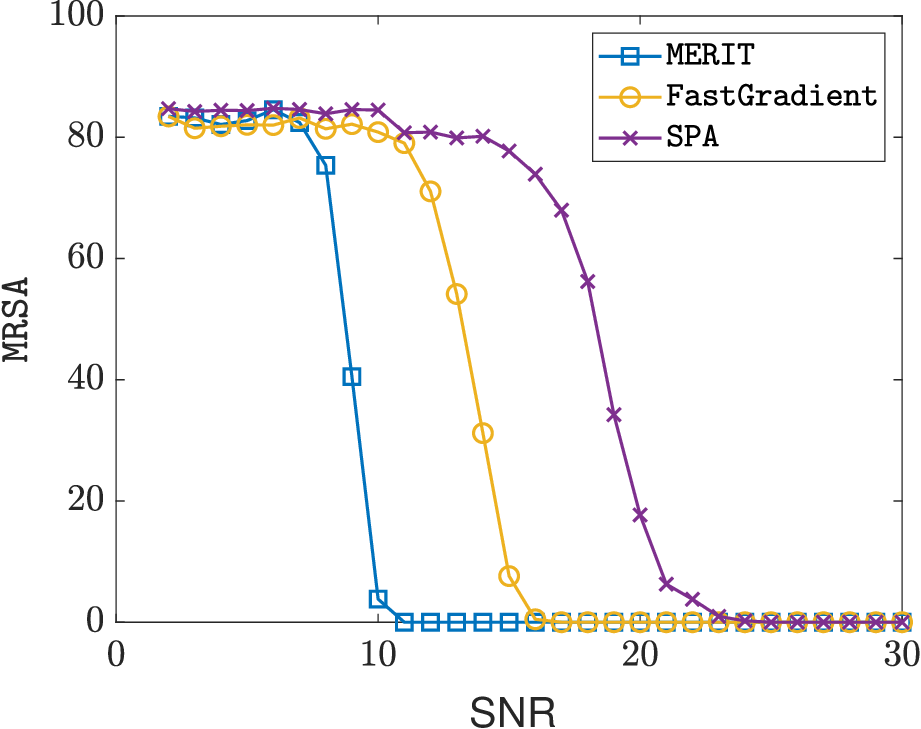}
    }
    \subfloat[RAE]{
        \includegraphics[width=0.23\textwidth]{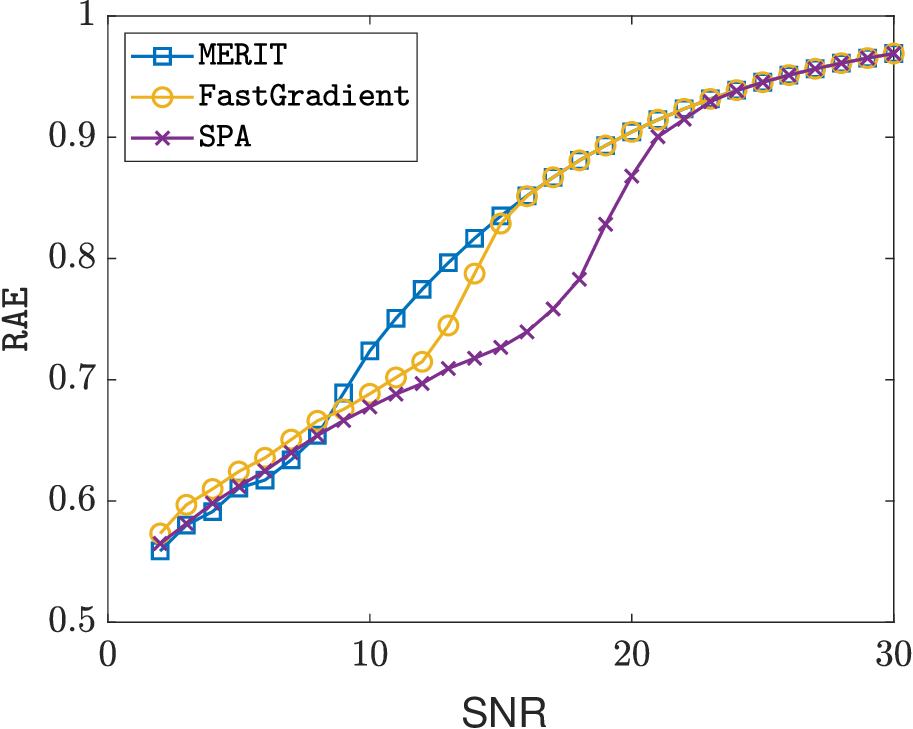}
    }
    \caption{The MRSAs and RAEs of the algorithms under various SNRs. $(M,N,K)=(50,55,10)$.}\label{fig:MRSARAE}
\end{figure}

Fig.~\ref{fig:ratenew} shows the success rates of different methods under the generative model and setting from \cite{gillis2018afast}, where $(M,N,K)=(50,55,10)$. 
One can see that \texttt{FastGradient} and \texttt{MERIT} both exhibit more satisfactory performance relative to \texttt{SPA}, which echos our comment that the all-at-once convex approaches often have better noise robustness relative to greedy pursuit. In particular, \texttt{FastGradient} and \texttt{MERIT} reach 100\% success rates at SNR$=16$dB and SNR$=10$dB, respectively, while \texttt{SPA} does not reach this accuracy even when SNR$=20$dB.
Fig.~\ref{fig:MRSARAE} shows the MRSAs and RAEs of the algorithms under the same setting, where similar observations are made.

\begin{figure}[t]
    \centering
    \captionsetup[subfigure]{margin=10pt}
    \subfloat[Success rate under various SNRs; $N=200$.  ] {
    \includegraphics[width=.44\linewidth]{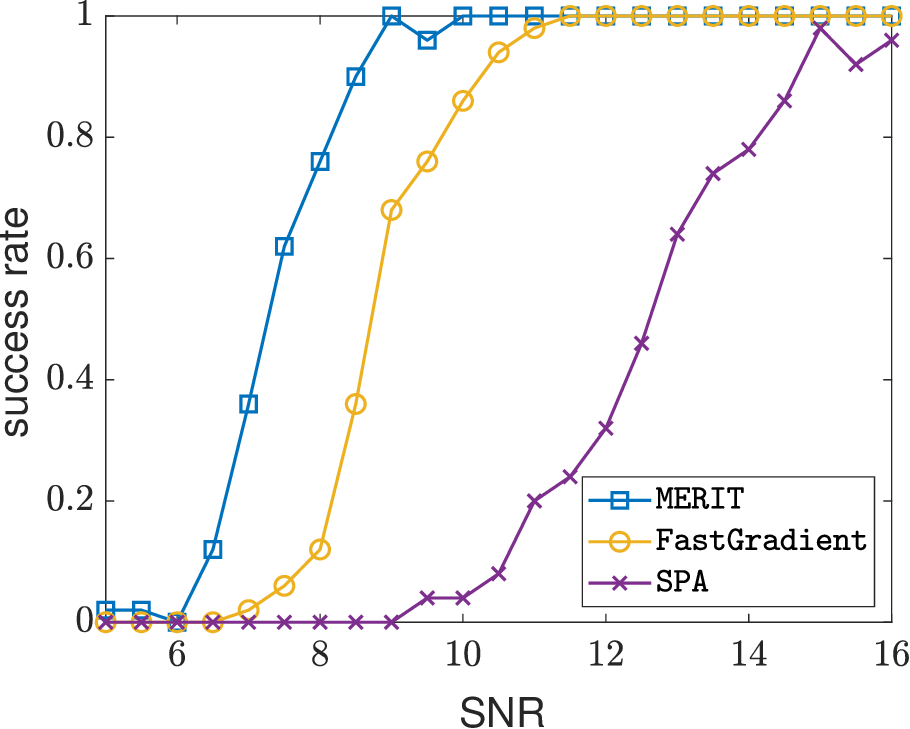}
    }
    \subfloat[Memory consumption under various $N$s; SNR=$10$dB.]{
        \label{fig:synthetic-memory}
        \includegraphics[width=.45\linewidth]{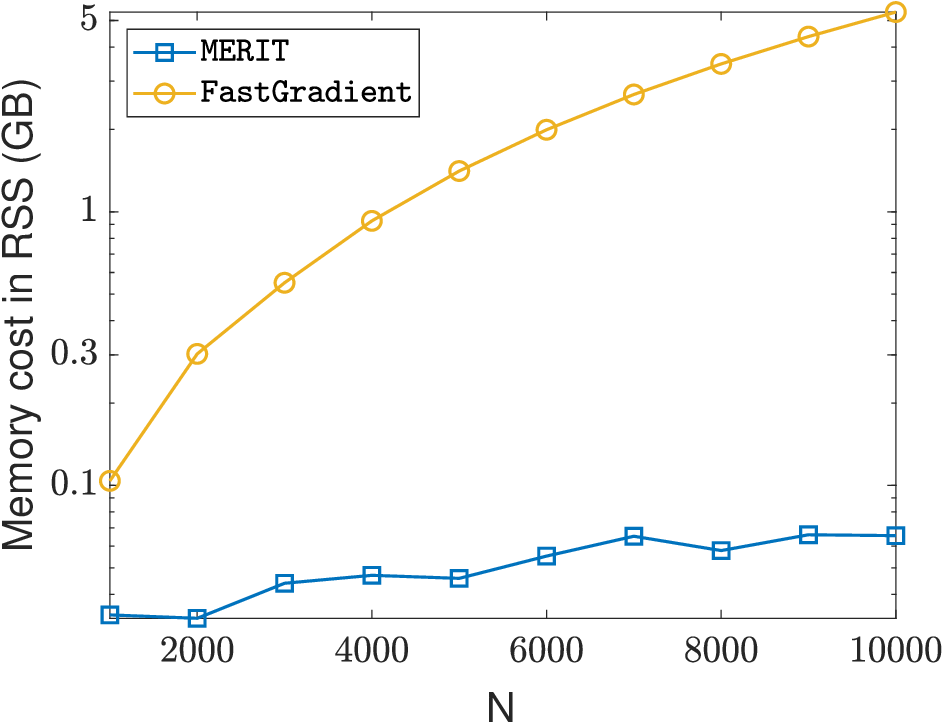}
    }
    \caption{Success rate performance and memory costs of the algorithms; $M=50,K=40$.}  \label{fig:syn_succ_rate}
\end{figure}

Fig.~\ref{fig:syn_succ_rate} (a) shows the success rates of the algorithms under different SNRs using $(M,K,N)=(50,40,200)$
and $\H(:,{\cal K}^c)$'s that are less special than that in the previous case; i.e., $\H(:,{\cal K}^c)$'s columns are generated following the uniform Dirichlet distribution with its parameter being $\bm 1\in\mathbb{R}^K$.
One can see that the algorithms perform similarly as in the case of Fig.~\ref{fig:ratenew}, except that the gap between \texttt{SPA} and the convex approaches \texttt{MERIT} (with regularization) and \texttt{FastGradient} is larger than that in the previous case.

\begin{table}[t]
    \centering
    \caption{Performance of the algorithms under various $K$'s; $N=200, M=80$, SNR=10dB.}
    \resizebox{\linewidth}{!}{\Large
    \begin{tabular}{|c|c|c|c|c|c|c|c|c|c|}
        \hline
        \multirow{2}{*}{$K$} & \multicolumn{3}{|c|}{success rate} & \multicolumn{3}{|c|}{MRSA} &  \multicolumn{3}{|c|}{RAE} \\ 
        \cline{2-10} 
        & \texttt{SPA} & \texttt{FastGradient} & \texttt{MERIT} &
         \texttt{SPA} & \texttt{FastGradient} & \texttt{MERIT} &
         \texttt{SPA} & \texttt{FastGradient} & \texttt{MERIT} \\
         \hline
         40 &0.98 & 0.98 & \textbf{1.00}  &54.7824 &54.7724 &\textbf{54.7292} &\textbf{0.7686} &\textbf{0.7686} & \textbf{0.7686}\\
         50 &0.84 & \textbf{1.00} & \textbf{1.00}  &55.4776 &\textbf{55.1517} &\textbf{55.1517} &0.7827 &\textbf{0.7830} & \textbf{0.7830}\\
         60 &0.42 & \textbf{1.00} & \textbf{1.00}  &56.4556 &\textbf{55.2206} &\textbf{55.2206} &0.7941 &\textbf{0.7955} & \textbf{0.7955}\\
         70 &0.00 & \textbf{1.00} & \textbf{1.00}  &58.9853 &\textbf{55.6658} &\textbf{55.6658} &0.8022 &\textbf{0.8069} & \textbf{0.8069}\\
         \hline
    \end{tabular}}\label{tbl:small_M}
\end{table}

Fig.~\ref{fig:syn_succ_rate} {(b)}  shows the memory costs of the two convex optimization-based algorithms under different $N$'s. Here, we set ${(M,K)=(50, 40)}$ and SNR$=10$dB. The memory is measured in terms of  maximum {\it resident set size} (RSS), which is the amount of allocated memory in RAM for a running program. The RSS is measured using a Linux built-in command named \texttt{time}\footnote{https://man7.org/linux/man-pages/man1/time.1.html}. 
One can see that \texttt{MERIT}'s memory growth along with $N$ is very graceful, but \texttt{FastGradient} quickly reaches the level that is close to {5GB} when ${N=10,000}$---while \texttt{MERIT} uses less than 0.1GB memory under the same problem size. 

Table~\ref{tbl:small_M} shows the performance of the algorithms under various $K$'s. 
As expected, \texttt{SPA} works better when $K$ is relatively small. 
The performance deteriorates when $K$ increases, showing the effect of error accumulation.
The convex approaches work similarly and exhibit consistently good performance across all the $K$'s under test.

\color{black}

\begin{figure}[t!]
    \centering
    \subfloat[Varying $\lambda$; $\mu = 1\mathrm{e}{-5}$.]{
        \includegraphics[width=0.22\textwidth]{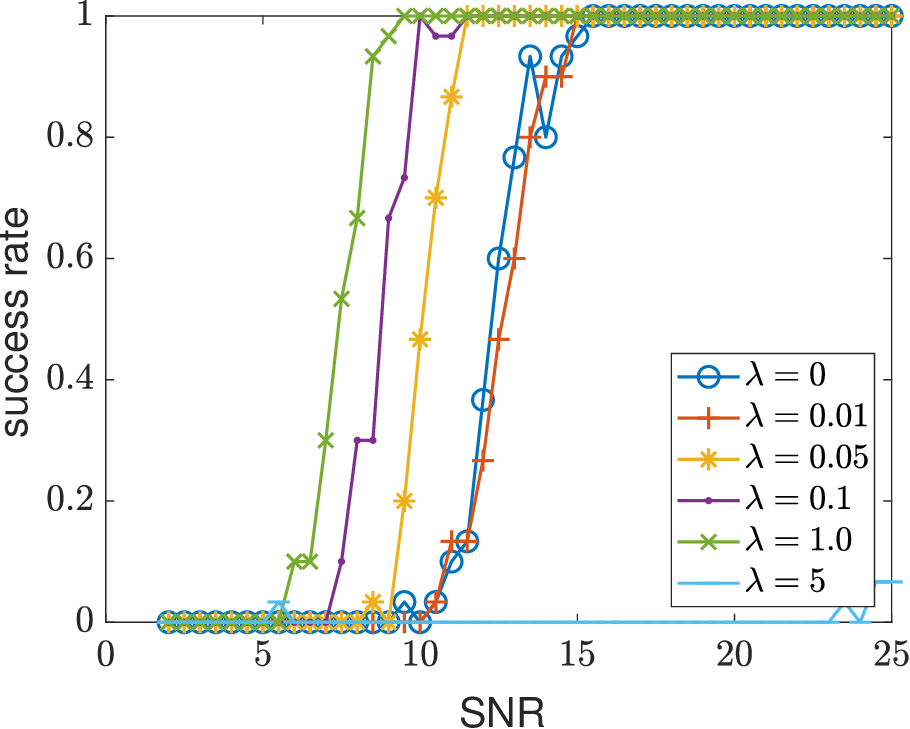}
    }
    \subfloat[Varying $\mu$; $\lambda = 0.1$.]{
        \includegraphics[width=0.22\textwidth]{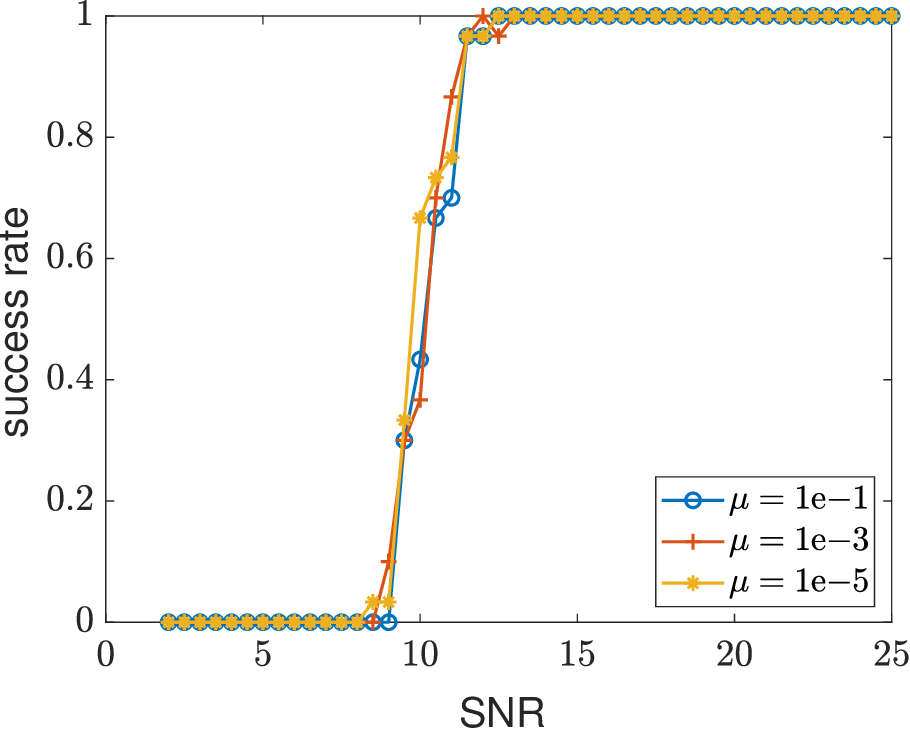}
    }
    \centering
    \caption{{Performance varies on different $\lambda$ and $\mu$; results are averaged over $30$ trials; $(M, K, N) =  (50, 40, 200)$}.}
    \label{fig:sensitivity}
\end{figure}

Fig.~\ref{fig:sensitivity} shows the impact of the hyperparameters $\lambda$ and $\mu$ on the \texttt{MERIT} algorithm. One can see that for lower SNRs, a larger $\lambda$ often works better---which is consistent with our analysis and intuition. The parameter $\mu$ is less consequential. That is, the wide range of $\mu$'s tested in our simulations give almost the same success rate curves.

\subsection{Real Data Experiment: Topic Modeling}
\noindent
{\bf Data.} We use two popular datasets, namely, the NIST Topic Detection and Tracking (TDT2) and the Reuters-21578 corpora, for the evaluation. 
Following the settings in topic modeling papers, e.g., \cite{cai2010locally,huang2016anchor}, we use single-topic documents so that
classic evaluation metrics (e.g., clustering accuracy \cite{cai2010locally,cai2005document}) can be easily used.
TDT2 contains ${D=8,384}$ single-topic documents with ${N=36,771}$ words as its vocabulary, while Reuters-21578 contains ${D=8,293}$ single-topic document with  ${N=18,933}$ words in its vocabulary.
All stop words are removed before running each trial. 

Note that we follow the formulation in \cite{arora2012practical,fu2018anchor,huang2016anchor} that applies NMF in the correlation domain. That is, the word-word correlation (or co-occurrence) data matrix $\X$ has a size of $N\times N$. The matrix $\X$ is estimated by the Gram matrix of the word-document term-frequency-inverse-document-frequency (TF-IDF) representation of the data; see more details in \cite{fu2018anchor,huang2016anchor}. Following the work in \cite{fu2014self}, a pre-processing step for noise reduction is used in this subsection. Specifically, before running the separable NMF algorithms, the principal component analysis (PCA)-based dimensionality reduction (DR) is used to reduce the row dimension of the co-occurrence matrix to $M'=2K$, which serves as an over estimate for $K$. In practice, such DR method can be easily implemented with any $M'>K$ if $K$ is roughly known. After the DR process, one factor of the dimension-reduced factorization model may not be nonnegative. Thus, technically, the model is not ``NMF''. Nonetheless, all the SD-MMV methods can still be applied since the nonnegativity of the left factor ($\W$) is never used in attaining ${\cal K}$.

\noindent
{\bf Baseline and Algorithm Settings.} We use a number of separable NMF based topic modeling algorithms as our baselines in this experiment.
In addition to \texttt{SPA} and \texttt{FastGradient}, we also use \texttt{XRAY} \cite{kumar2013fast} and \texttt{FastAnchor} \cite{arora2012practical} that are both greedy algorithms and variants of \texttt{SPA} developed under the context of topic modeling.
The \texttt{LDA} \cite{blei2003latent} method using Gibbs sampling \cite{blei2012probabilistic} is also employed, as a standard baseline. 

We use warm start for both \texttt{MERIT} and \texttt{FastGradient}. Particularly, we use \texttt{SPA} to extract an estimate of ${\cal K}$, denoted as $\widehat{\mathcal{K}}_{\texttt{SPA}}$.
Then, we compute $\C^{\rm init}$ by $$\bm{C}^{\rm init}(\mathcal{K}_{\texttt{SPA}}, :) = \argmin_{\bm 1^\T\bm{H}=\bm 1^\T,\bm H\geq \bm 0} \|\bm{X} - \bm{X}(:, \widehat{\mathcal{K}}) \bm{H}\|_{\rm F}^2, 
$$ and let $\bm{C}^{\rm init}(\widehat{\mathcal{K}}_{\texttt{SPA}}^c, :) = \bm{0}.
$
For \texttt{MERIT}, we set $\lambda=10^{-6}$ and $\mu=10^{-5}$ in all cases. We try multiple $\lambda$'s for \texttt{FastGradient} and present the best performance it attains. For \texttt{MERIT}, since we use WS-FW (cf. Sec.~\ref{sec:prelim}), the $t_{\rm init}$ needs to be determined. In the WS-FW paper \cite{freund2016new},  $t_{\rm init}$ in theory is computed using the curvature constant associated with optimization problem and initial duality gap. However, as admitted in \cite{freund2016new}, estimating the curvature constant is still an open challenge. Hence, we use a heuristic that $t_{\rm init}={\rm round}(1/{\rm RMSE}_{\rm init})$, where ${\rm RMSE}_{\rm init}=\sqrt{\nicefrac{1}{N} \|\bm{X} - \bm{X}\bm{C}^{\rm init}\|_{\rm F}^2}$. This reflects the idea that a better initialization should use a larger $t_{\rm init}$.

\noindent
{\bf Metrics.} We use the three metrics from \cite{huang2016anchor,fu2018anchor}, namely, coherence (\texttt{Coh}), word similarity count (\texttt{SimCount}), and clustering accuracy (\texttt{ClustAcc}).  The \texttt{Coh} metric evaluates if a single topic's quality by measuring the coherence of the high-frequency words contained in this topic. The \texttt{SimCount} metric measures how diverse are the mined topics. The \texttt{ClustAcc} compares the estimated $\widehat{\bm H}$ with the ground-truth labels after automatic permutation removal using the Kuhn-Munkres algorithm; see more details of the evaluation process in \cite{fu2018nonnegative,huang2016anchor} and the definition of \texttt{ClustAcc} in \cite[Sec. 5.2]{cai2005document}..
A good topic mining algorithm is expected to attain high \texttt{Coh} values, low \texttt{SimCount} and high \texttt{ClustAcc} values.
Among the three, \texttt{ClustAcc} is arguably the most indicative for the quality of the mined topics if the objective is to use the topics for downstream tasks. 
For each trial, we randomly draw documents associated with $K$ topics from the datasets and apply the algorithms. The results for each $K$ are averaged from 50 trials.

\noindent
{\bf Results}.
Tables \ref{table:tdt2_performance} shows the performance of the algorithms on TDT2 and Reuters-21578, respectively. One can see
that both the regularized and unregularized versions of the proposed method, i.e., \texttt{MERIT} and \texttt{MERIT(0)} (i.e., the version of \texttt{MERIT} with $\lambda=0$), exhibit competitive performance.
The proposed methods outperform all the baselines on TDT2 in terms of \texttt{Coh} and \texttt{ClustAcc}.
The \texttt{SimCount} performance of the proposed algorithms is also reasonable.
In particular, when $K=10$, the \texttt{ClustAcc} of \texttt{MERIT} exhibits a 5\% improvement compared to the best baseline, which is a remarkable margin.
On Reuters-21578, the \texttt{MERIT} method also consistently offers the best and second best performance in terms of \texttt{ClustAcc} in most cases.

From these results, one can see that the convex optimization based separable NMF algorithms, i.e., \texttt{FastGradient}, \texttt{MERIT} and \texttt{MERIT(0)}, in general work better than the greedy methods, namely, \texttt{SPA}, \texttt{XRAY}, and \texttt{FastAnchor}. This is consistent with our observation in the synthetic data experiments. This advocates using such all-at-once algorithms for real applications.
Another observation is that \texttt{MERIT} slightly outperforms \texttt{MERIT(0)}, which shows that the designed regularization is still effective on topic modeling problems.

\begin{table*}[t]
    \fontsize{7.}{8.4}\selectfont
    \setlength\tabcolsep{2pt} 
    \caption{Performance of the algorithms on: (Left) TDT2; vocabulary size$=36,771$, and (Right) Reuters-21578; vocabulary size = $18,933$}
    \label{table:tdt2_performance}
    \begin{tabular}{|p{1.2cm}|c|c|c|c|c|c|c|c|c|}
        \multicolumn{10}{c}{TDT2} \\
        \hline
        & \textbf{Method / $K$} & $3$ & $4$ & $5$ & $6$ & $7$ & $8$ & $9$ & $10$ \\
        \hline
        \multirow{6}{*}{Coherence} 
        & \texttt{SPA} & {\blue -346.6} & -388.4 & -404.9 & -432.0 & -418.6 & -438.2 & -443.5 & -456.7 \\
        \cline{2-10}
        & \texttt{FastAnchor} & -468.6 & -483.4 & -483.3 & -495.9 & -525.8 & -536.2 & -546.5 & -543.2 \\
        \cline{2-10}
        & \texttt{XRAY} & -347.4 & -389.2 & -405.4 & -432.0 & -419.0 & -439.4 & -443.2 & -459.2 \\
        \cline{2-10}
        & \texttt{LDA} & -521.6 & -526.2 & -530.4 & -546.0 & -550.0 & -538.8 & -543.1 & -553.1 \\
        \cline{2-10}
        & \texttt{FastGradient} & -553.8 & -517.1 & -537.2 & -534.6 & -561.9 & -562.7 & -571.9 & -585.5 \\
        \cline{2-10}
          & \texttt{MERIT} & -351.5 & \textbf{-375.7} & \textbf{-385.8} & \textbf{-394.4} & \textbf{-399.3} & \textbf{-417.2} & \textbf{-417.5} & \textbf{-429.1} \\
        \cline{2-10}
          & \texttt{MERIT}$(0)$ & \textbf{-345.0} & {\blue -388.4} & {\blue -404.8} & -433.4 & -420.1 & -439.4 & -444.3 & -458.3  \\
        \hhline{|=|=|=|=|=|=|=|=|=|=|}
        \multirow{6}{1.6cm}{Similarity Count}
        & \texttt{SPA} & {\blue 1.06} & 3.64   & 5.76   & 10.24  & 14.24  & 23.18  & 27.56  & 43.62  \\
        \cline{2-10}
        & \texttt{FastAnchor} & {\blue 1.06} & \textbf{2.02} & \textbf{3.90} & \textbf{4.80} & \textbf{6.18} & \textbf{7.98} & \textbf{9.92} & \textbf{11.22} \\
        \cline{2-10}
        & \texttt{XRAY} & \textbf{1.00} & 3.88 & 5.66 & 10.24& 14.16& 23.18& 28.00& 43.4 \\
        \cline{2-10}
        & \texttt{LDA} & 1.08 & {\blue 2.96} & {\blue 5.62} & {\blue 7.84} & {\blue 12.24}& {\blue 17.28}& {\blue 21.84}& {\blue 27.5} \\
        \cline{2-10}
        & \texttt{FastGradient} & 14.80 & 26.34 & 47.16 & 62.28 & 71.24 & 100.58& 109.84& 127.32 \\
        \cline{2-10}
        & \texttt{MERIT} & 1.56 & 4.98 & 5.76 & 7.92 & 13.30 & 21.16 & 28.52 & 36.08 \\
        \cline{2-10}
          & \texttt{MERIT}$(0)$ & 1.06 & 3.64 & 5.78 & 10.56 & 14.38 & 22.62 & 27.50 & 43.06 \\
        \hhline{|=|=|=|=|=|=|=|=|=|=|}
        \multirow{6}{*}{Accuracy}
          & \texttt{SPA} & {\blue 0.87} & {\blue 0.83} & {\blue 0.81} & {\blue 0.81} & {\blue 0.78} & {\blue 0.76} & {\blue 0.75} & {\blue 0.72} \\
        \cline{2-10}
        & \texttt{FastAnchor} & 0.77 & 0.72 & 0.67 & 0.63 & 0.66 & 0.63 & 0.65 & 0.65  \\
        \cline{2-10}
        & \texttt{XRAY} & {\blue 0.87} & 0.82 & 0.80 & {\blue 0.81} & {\blue 0.78} & {\blue 0.75} & {\blue 0.75} & 0.71 \\
        \cline{2-10}
        & \texttt{LDA} & 0.78 & 0.77 & 0.74 & 0.75 & 0.73 & 0.72 & 0.68 & 0.70 \\
        \cline{2-10}
        & \texttt{FastGradient} & 0.70 & 0.71 & 0.65 & 0.64 & 0.61 & 0.56 & 0.58 & 0.57 \\
        \cline{2-10}
          & \texttt{MERIT} & \textbf{0.88} & \textbf{0.88} & \textbf{0.85} & \textbf{0.86} & \textbf{0.84} & \textbf{0.82} & \textbf{0.80} & \textbf{0.77} \\
        \cline{2-10}
          & \texttt{MERIT}$(0)$ & 0.86 & {\blue 0.83} & 0.80 & {\blue 0.81} & {\blue 0.78} & {\blue 0.76} & {\blue 0.75} & {\blue 0.72} \\
        \hline 
    \end{tabular}
    \;
    \begin{tabular}{|c|c|c|c|c|c|c|c|c|c|}
        \multicolumn{10}{c}{Reuters-21578} \\
        \hline
        & \textbf{Method / $K$} & $3$ & $4$ & $5$ & $6$ & $7$ & $8$ & $9$ & $10$ \\
        \hline
        & \texttt{SPA} & {\blue -402.7} & -416.4 & \textbf{-420.5} & -442.1 & -516.5 & -520.3 & {\blue -541.5} & {\blue -548.3} \\
        \cline{2-10}
        & \texttt{FastAnchor} & -655.0 & -681.0 & -693.6 & -711.1 & - 757.5 & -827.7 & -832.8 & -843.4 \\
        \cline{2-10}
        & \texttt{XRAY} & -404.4 & {\blue -415.2} & -422.7 & {\blue -441.6} & {\blue -516.3} & {\blue -519.6} & -542.2 & -548.6 \\
        \cline{2-10}
        & \texttt{LDA} & -674.1 & -677.2 & -686.3 & -715.2 & -705.9 & -762.9 & -776.8 & -776.5 \\
        \cline{2-10}
        & \texttt{FastGradient} & -657.1 & -768.3 & -782.0 & -821.8 & -847.1 & -967.7 & -989.5 & -959.8 \\
        \cline{2-10}
          & \texttt{MERIT} & -430.6 & -452.8 & -466.4 & -494.0 & -539.2 & -541.1 & -564.8 & -570.8 \\
        \cline{2-10}
          & \texttt{MERIT}$(0)$ & \textbf{-401.7} & \textbf{-413.3} & {\blue -422.5} & \textbf{-440.8} & \textbf{-511.2} & \textbf{-518.2} & \textbf{-536.0} & \textbf{-544.0} \\
        \hhline{|=|=|=|=|=|=|=|=|=|=|}
        & \texttt{SPA} & 7.46 & 15.16 & 23.82 & 51.98 & 59.38 & 158.50 & 235.62 & 219.16 \\
        \cline{2-10}
        & \texttt{FastAnchor} & \textbf{5.40} & {\blue 8.46} & {\blue 13.06} & {\blue 20.06} & {\blue 25.56} & {\blue 42.28} & {\blue 54.9} & {\blue 57.84} \\
        \cline{2-10}
        & \texttt{XRAY} & 6.76 & 14.18 & 23.82 & 52.06 & 59.64 & 160.96 & 235.10 & 221.50 \\
        \cline{2-10}
        & \texttt{LDA} & \textbf{3.20} & \textbf{6.46} & \textbf{9.32} & \textbf{12.48} & \textbf{21.22} & \textbf{24.60} & \textbf{33.56} & \textbf{39.68} \\
        \cline{2-10}
        & \texttt{FastGradient} & 12.96  & 20.62  & 30.42  & 47.56  & 60.46  & 82.86  & 106.66 & 144.38 \\
        \cline{2-10}
          & \texttt{MERIT} & 7.34 & 16.04 & 21.88 & 36.08 & 48.36 & 93.32 & 131.62 & 141.42 \\
        \cline{2-10}
          & \texttt{MERIT}$(0)$ & 7.38 & 15.18 & 23.24 & 45.12 & 54.60 & 145.52 & 223.66 & 214.60 \\
        \hhline{|=|=|=|=|=|=|=|=|=|=|}
          & \texttt{SPA} & {\blue 0.64} & 0.57 & {\blue 0.54} & 0.51 & 0.49 & 0.44 & 0.42 & 0.40 \\
        \cline{2-10}
        & \texttt{FastAnchor} & 0.60 & 0.57 & 0.52 & {\blue 0.52} & 0.46 & 0.42 & 0.38 & 0.37 \\
        \cline{2-10}
        & \texttt{XRAY} & 0.63 & 0.57 & {\blue 0.54} & 0.51 & 0.49 & {\blue 0.45} & 0.42 & 0.40 \\
        \cline{2-10}
        & \texttt{LDA} & 0.63 & 0.57 & 0.53 & 0.51 & 0.46 & 0.44 & 0.41 & 0.42 \\
        \cline{2-10}
        & \texttt{FastGradient} & 0.62 & 0.57 & \textbf{0.56} & 0.51 & {\blue 0.50} & \textbf{0.48} & \textbf{0.44} & \textbf{0.46} \\
        \cline{2-10}
          & \texttt{MERIT} & \textbf{0.66} & \textbf{0.62} & 0.53 & \textbf{0.53} & \textbf{0.51} & \textbf{0.48} & {\blue 0.43} & {\blue 0.45} \\
        \cline{2-10}
          & \texttt{MERIT}$(0)$ & {\blue 0.64} & {\blue 0.58} & {\blue 0.54} & {\blue 0.52} & 0.49 & 0.44 & 0.42 & 0.41 \\
        \hline 
    \end{tabular}
\end{table*}

\begin{figure}[t]
    \centering
    \subfloat[TDT2] {
        \includegraphics[width=0.22\textwidth]{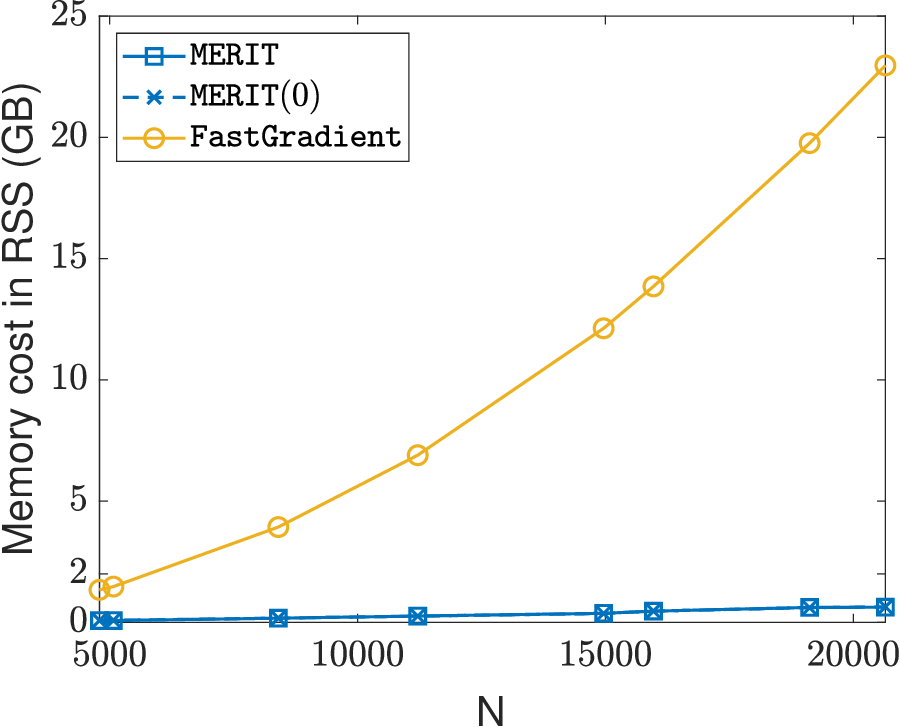}
        \label{fig:topic_mem_tdt2}
    }
    \subfloat[Reuters-21578] {
        \includegraphics[width=0.22\textwidth]{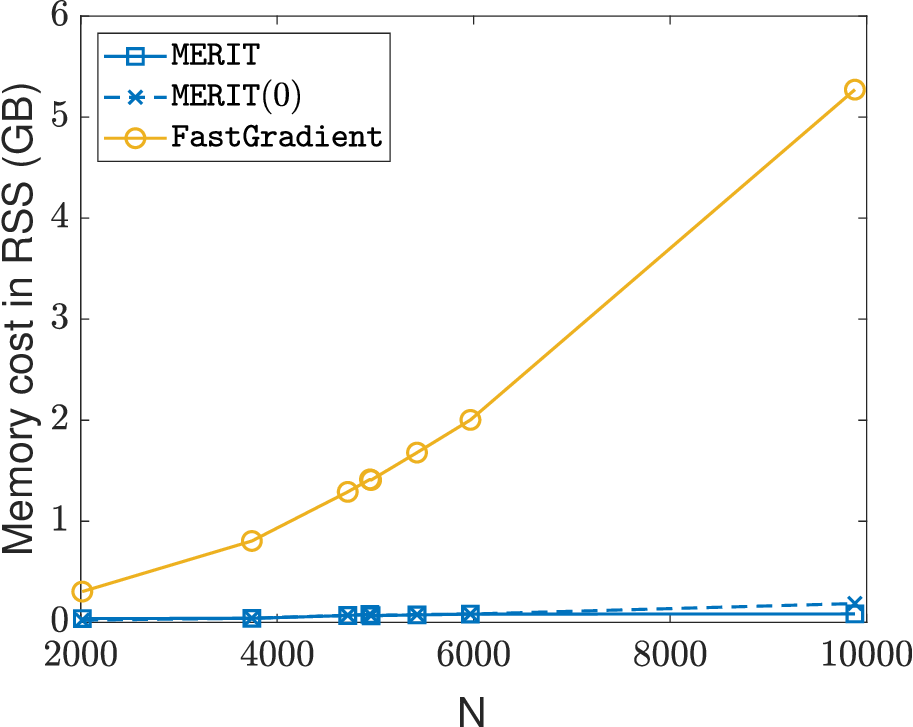}
        \label{fig:topic_mem_reuters}
    }
    \caption{Memory consumption of \texttt{FastGradient} and \texttt{MERIT}, under different sample sizes.}
    \label{fig:topic_mem}
\end{figure}

Fig.~\ref{fig:topic_mem} shows the memory consumption of the algorithms on TDT2 and Reuters-21578, respectively. 
Since the data matrix's size changes in each trial due to the varying stop words, we only plot the first trial for each $K$. 
One can see that when $N$ reaches $20,000$, \texttt{FastGradient} uses more than 20GB memory, while \texttt{MERIT} and \texttt{MERIT(0)} use less than 2GB.
We observe that \texttt{MERIT(0)} works well in the topic modeling experiment, perhaps because the data model is reasonably well aligned with that in \eqref{eq:nmf_noise} without much modeling error.

\subsection{Real Data Experiment: Community Detection}
The link between mixed membership stochastic model (MMSB)-based overlapped community detection and simplex-structured matrix factorization has often been used in the literature; see \cite{panov2018consistent,mao2017mixed,huang2019detecting,ibrahim2020mixed}. By applying eigendecomposition to a node-node undirected adjacency matrix, the membership estimation problem boils down to a noisy matrix factorization under probability simplex constraints. 
This leads to $\X\approx \W\H$, where $\X\in\mathbb{R}^{K\times N}$ is extracted by the eigendecomposition of the $N\times N$ adjacency matrix (the rows of $\X$ are the first $K$ eigenvectors), $\H$'s columns are the mixed membership vectors correspond to the $N$ node. More precisely, $h_{k,\ell}$ is the probability that node $\ell$ is associated with community $k$. This physical interpretation also means that $\bm 1^\T\bm H=\bm 1^\T$ and $\bm H\geq \bm 0$.
Hence, both \texttt{SPA} and convex optimization based separable NMF algorithms can be applied to tackle this problem, if the separability condition holds. In the context of community detection, separability is equivalent to the existence of ``pure nodes'' for each community, i.e., nodes who are only associated with a single community.

\noindent
{\bf Data.} We use two co-authorship networks, namely, the Data Base systems and Logic Programming (DBLP) data and the Microsoft Academic Graph (MAG) data. A community in DBLP is defined as a group of conferences. The ``field of study'' is considered as a community in MAG. The ground-truth memberships of the nodes in these two datasets are known;  see the \texttt{Matlab} format of this data from \cite{mao2017mixed}.

In our experiments, we consider the nodes who contribute to 99\% of the energy (in terms of squared Euclidean norm).
The remaining nodes are not used in the algorithms due to their rare collaboration with others.
The detailed statistics of the used DBLP and MAG data are given in Table~\ref{table:comm_stats}.

\begin{table}[t]
    \centering
    \caption{Statistics of the DBLP and MAG datasets Used In The Experiments.}
    \label{table:comm_stats}
    \begin{tabular}{|c | c|c|}
        \hline
        \textbf{Dataset  }&  \textbf{\# nodes }& \textbf{\# communities} \\  \hline \hline
        DBLP1 & 6437 & 6  \\ \hline
        DBLP2 & 3248&  3 \\ \hline
        DBLP3 & 3129&  3 \\ \hline
        DBLP4 & 1032&  3 \\ \hline
        DBLP5 & 3509&  4 \\ \hline
        MAG1  &  1447& 3  \\ \hline
        MAG2  &  4974& 3  \\ \hline
    \end{tabular}
\end{table}
\noindent
{\bf Baselines.} 
 We compare \texttt{MERIT} and \texttt{MERIT(0)} with three baselines. The first two algorithms are \texttt{GeoNMF} \cite{mao2017mixed} and \texttt{SPOC} \cite{panov2018consistent} as they are reportedly popular within the class of greedy method in context of community detection. And \texttt{FastGradient} as a candidate for the convex optimization based approach is included.
\texttt{MERIT} uses the same hyperparameters setting as in the topic modeling problem, i.e., ${\lambda = 10^{-6},\mu = 10^{-5}}$. Again, we try multiple $\lambda$'s for \texttt{FastGradient} and report its best performance.

\noindent
{\bf Metric.} Following \cite{mao2017mixed,panov2018consistent,ibrahim2020mixed}, we evaluate the performance based on the {\it averaged Spearman's rank correlation} (SRC) coefficient between the learned community membership matrix $\widehat{\H}$ and the ground-truth $\bm{H}$. 

By the definition, the \texttt{SRC} measures the ranking similarity between the estimated and ground-truth mixed membership vectors. The estimated $ \widehat{\bm{H}}$ is obtained by probability simplex-constrained least squares using data $\X$ and the basis $\widehat{\W}=\X(:,{\cal \widehat{K}})$. 
By definition, SRC can take value in the interval from $-1$ to  $1$. A higher value indicates a better alignment between  $\widehat{\bm{H}}$ and $\bm{H}$.

\begin{table}[t]
    \centering
    \caption{SRC Performance on DBLP and MAG.}
        \label{table:comm_result}
        \fontsize{8}{9.6}\selectfont
        \setlength\tabcolsep{3pt} 
        \begin{tabular}{|c|c|c|c|c|c|}
            \hline
            \textbf{Dataset}   &   \texttt{GeoNMF} & \texttt{SPOC} &  \texttt{FastGradient} & \texttt{MERIT}& \texttt{MERIT(0)}   \\  \hline \hline
            DBLP1 &  0.2974 & {\blue 0.2996} &  \textbf{0.3145} & 0.2937 & 0.2912 \\
            DBLP2 &  0.2948 & 0.2126 &  {\blue 0.3237} & \textbf{0.3257} & 0.2931 \\
            DBLP3 &  0.2629 & \textbf{0.2972} &  0.1933 & 0.2763 & {\blue 0.2766} \\
            DBLP4 &  0.2661 & {0.3479} &  0.1601 & \textbf{0.3559} & \textbf{0.3559} \\
            DBLP5 &  {\blue 0.1977} & 0.1720 &  0.0912 & \textbf{0.1983} & \textbf{0.1983} \\
            MAG1  &  \textbf{0.1349} & {\blue 0.1173} &  0.0441 & 0.1149 & 0.1074 \\
            MAG2  & 0.1451& 0.1531 & {\bf 0.2426} & {\blue 0.2414} & 0.1374 \\
            \hline
        \end{tabular}

\end{table}
       
\begin{figure}
    \centering
    \includegraphics[width=0.8\linewidth]{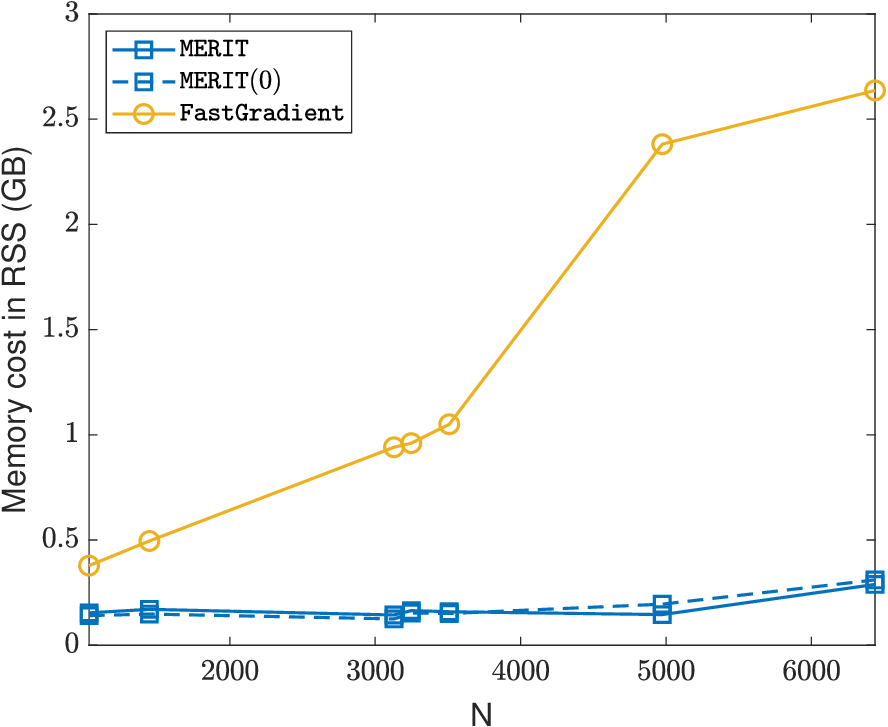}
    \caption{The memory consumption of \texttt{MERIT} and \texttt{FastGradient} after the eigendecomposition step.}
    \label{fig:mem_commdetct}
\end{figure}

\noindent
{\bf Results.} From Table~\ref{table:comm_result}, one can see that the proposed \texttt{MERIT} method offers competitive SRC results over all the datasets under test. 
Similar as before, the convex all-at-once algorithms, especially, \texttt{MERIT}, and \texttt{MERIT(0)}, often output competitive SRC results.
In particular, the \texttt{MERIT} class' SRC values are in top 2 over five out of seven datasets. 
The greedy methods (i.e., \texttt{GeoNMF}, and \texttt{SPOC}) also work reasonably well, but less competitive in some datasets. For example, in MAG2 and DBLP2, the performance gap between the greedy algorithms and the convex methods are particularly articulate.

Fig.~\ref{fig:mem_commdetct} compares the memory consumption of \texttt{FastGradient} and \texttt{MERIT}. 
 As  $N$ reaches  $6500$, \texttt{FastGradient} consumes more than $2$GB while \texttt {MERIT} and \texttt {MERIT(0)} use significant less memory, i.e., under $0.3$GB. This is consistent with our observations in the topic modeling examples.
 
\smallskip

A final remark is that the theorems in this work are based on worst-case analyses,
and thus the noise bounds and identification conditions are naturally pessimistic. However, note that our conditions are only sufficient (instead of sufficient and necessary). 
This may explain the reason why the proposed method works under many settings of the experiments where the noise level is quite high.

\section{Conclusion}%
\label{sec:conclusion}
In this work, we revisited convex optimization-based self-dictionary SD-MMV for separable NMF. 
This line of work emerged about a decade ago as a tractable and robust means for solving the challenging NMF problem.
The method is recognized as an important category of NMF algorithms, but has serious challenges when dealing with big data. In particular, existing convex SD-MMV approaches' memory complexity scales quadratically with the number of samples, which is hardly feasible for datasets that have more than 1000 samples.
We proposed a new algorithm based on the Frank-Wolfe method, or, conditional gradient. 
Unlike existing algorithms, our method is shown to have linear memory complexity under mild conditions---even in the presence of noise. For performance enhancement, we also offered a smoothed row-sparsity-promoting regularizer, and showed that it can provide stronger guarantees for memory efficiency against noise under the FW framework. We tested the algorithm using synthetic data and real-world topic modeling and community detection datasets. The results corroborate our theoretical analyses.

\bibliographystyle{IEEEtran}
\bibliography{refs}

\appendices

\section{Proof of Theorem \ref{theorem:noiseless}}\label{app:noiseless}
    Since $\norm{\bm{X} - \bm{X}\bm{C}}_{\rm F}^2 = \sum^{N}_{\ell=1} \norm{\bm{X}\bm{c}_{\ell} - \bm{x}_{\ell}}_2^2$, i.e., the cost function is decomposable over the columns of $\X$, one can consider each $f(\bm{c}_\ell) = \nicefrac{1}{2}\norm{\bm{X}\bm{c}_{\ell} - \bm{x}_{\ell}}_2^2$ individually.
    
    The FW algorithm first finds ${\bm u}$ such that $\nabla_{\bm c_\ell} f(\c_\ell)^\T\bm u$ is minimized
    over the probability simplex. This is a linear program, and its solution is always attained at a vertex of the probability simplex \cite{boyd2004convex}. Hence, the solution is an unit vector, i.e.,
\[
   \bm{u} =\bm{e}_{n^{\star}}, ~n^{\star} = \arg\min_{n\in [ N] } ~ [\nabla f(\mathbf{c}_\ell)]_n.
\] 

   Next, we show that $n^{\star} \in{\cal K}$ always holds.
To see this, we have
\begin{align*}
    \nabla f(\mathbf{c}_\ell) 
    &=\bm{X}^\T (\bm{X}\bm{c}_\ell -\bm{x}_\ell) \\
    &=\bm{H}^\T\bm{W}^\T\bm{W} (\H\bm{c}_\ell -\bm{h}_\ell)  \\
    &= \begin{bmatrix}
       \bm{h}_1^\T\bm{W}^\T\bm{W}(\H\bm{c}_\ell -\bm{h}_\ell) \\
    \vdots\\
       \bm{h}_N^\T\bm{W}^\T\bm{W}(\H\bm{c}_\ell -\bm{h}_\ell) \\
    \end{bmatrix} 
\end{align*}

Recall that $\bm{q}_\ell =\bm{W}^\T\bm{W} (\H\bm{c}_\ell -\bm{h}_\ell) \in \mathbb{R}^{K}$. Hence, we have
\begin{align*}
    [\nabla f(\bm{c}_\ell)]_n &=\bm{h}_n^\T\bm{q}_\ell \\
&    = \sum^{K}_{i=1} h_{i,n} q_{i,\ell}\\
    & \geq \left( \min_{j\in[ K]} q_{j,\ell } \right)  \sum^{K}_{j=1} h_{j,n} \\
    & = \min_{j\in[ K]} q_{j,\ell }   \numberthis \label{eq:noiseless_lowerbound}
\end{align*} 
If $\bm{q}_\ell \neq \bm{0}$, then the lower bound can be attained if $\bm{h}_n = \bm{e}_{j^{\star}}$ where $j^{\star}=\argmin_{j\in[K]} q_{j,\ell}$. 
Note that such $\bm{h}_n$ always exists because of the separability assumption. Let us denote it as $\bm{h}_{n^{\star}}$, we have $n^{\star} = \argmin_{n} [\nabla f(\bm{c}_\ell)]_n$. Since $\bm{h}_{n^{\star}}$ is a unit vector, and because of the assumption that non-repeated unit vectors appears in $\bm{H}$, one can conclude $n^{\star} \in \mathcal{K}$.

If there are more than one smallest element in $\bm{q}_\ell$, say, there are 2 smallest elements $q_{1,\ell}= q_{2,\ell} = \min_{j} {q}_{j,\ell}$, then, any $\bm{h}_{n^{\star}} $ in a form of $\bm{h}_{n^{\star}} = \begin{bmatrix}
            z,1- z, 0 ,\ldots , 0
        \end{bmatrix}^\T  $ for any $0\leq z \leq 1$ also makes the lower bound in \eqref{eq:noiseless_lowerbound} attained. Such $\bm{h}_{n^{\star}}$ might be not a unit vector and hence $n^{\star} \notin \mathcal{K}$. Nonetheless, the assumption that no duplicate minimal values appear in the gradient [cf. Eq.~\eqref{eq:nodupq}] assures that this case never happens.

If $\bm{q}_\ell =\bm{0} $, which means that the objective function already reaches optimal value of $0$ since 
        \[
            \bm{W}^\T\bm{W} (\H\bm c_\ell -\bm{h}_\ell) =\bm{0} \Longrightarrow\bm{H}\bm{c}_\ell = \bm{h}_\ell .
        \] 
Therefore, one should stop FW.

\section{Proof of Theorem \ref{theorem:noisy_unreg}}
\label{appedix:proof_theorem_noisy_unreg}

Under the noisy model,
for any $\bm c_\ell$, we have
\begin{align*}
    \nabla f(\bm{c}_\ell)
    &= \bm{X}^\T (\bm{X}\bm{c}_\ell - \bm{x}_\ell) \\
    &= (\bm{W}\bm{H} + \bm{V})^\T ((\bm{W}\bm{H} + \bm{V})\bm{c}_\ell - (\bm{W}\bm{h}_\ell + \bm{v}_\ell)). 
\end{align*}    
Hence, 
\begin{align*}
    & [\nabla f(\bm{c}_\ell)]_n 
    = (\bm{h}_n^\T \bm{W}^\T + \bm{v}_n^\T ) ((\bm{W}\bm{H} + \bm{V})\bm{c}_\ell - \bm{W}\bm{h}_\ell - \bm{v}_\ell) \\
    &= (\bm{h}_n^\T \bm{W}^\T + \bm{v}_n^\T ) (\bm{W}\bm{H}\bm{c}_\ell + \bm{V}\bm{c}_\ell - \bm{W}\bm{h}_\ell - \bm{v}_\ell) \\
    &= \bm{h}_n^\T \bm{W}^\T\bm{W} (\bm{H}\bm{c}_\ell - \bm{h}_\ell) \\
    &+ \underbrace{\bm{h}_n^\T \bm{W}^\T (\bm{V}\bm{c}_\ell - \bm{v}_\ell) + \bm{v}_n^\T (\bm{W}\bm{H}\bm{c}_\ell + \bm{V}\bm{c}_\ell -\bm{W}\bm{h}_\ell - \bm{v}_\ell)}_{\epsilon_n} \\
    &= \bm{h}_n^\T \bm{W}^\T \bm{W} (\bm{H}\bm{c}_\ell - \bm{h}_\ell) + \epsilon_n. \numberthis \label{eq:gradient_noise}
\end{align*}
Similar to the proof of Theorem~\ref{theorem:noiseless}, the key is to show that for $1 \leq \ell \leq N$, $n_\ell^\star \in {\cal K}$, where
\[
  n^\star_\ell =    \argmin_n \; \bm{h}_n^\T \bm{W}^\T \bm{W}(\bm{H}\bm{c}_\ell - \bm{h}_\ell) + \epsilon_n.
\] 
Since the proof in the sequel holds for all $\ell$, we omit the subscript and use $n^\star$ instead.
From the proof of Theorem~\ref{theorem:noiseless}, when $\epsilon_n = 0$, we know that $\bm{h}_{n^{\star}} = \bm{e}_j$ where $\bm{e}_{j}$ is a certain unit vector.
Our goal is to show that with such $n^{\star}$, for any $n \neq n^{\star}$, 
 \begin{align*}
    & \bm{h}_n^\T \bm{W}^\T \bm{W} (\bm{H}\bm{c}_\ell - \bm{h}_\ell) + \epsilon_n > \bm{h}_{n^{\star}}^\T \bm{W}^\T \bm{W} (\bm{H}\bm{c}_\ell - \bm{h}_\ell) + \epsilon_{n^{\star}}.
\end{align*}
Equivalently, we hope to show the following
\begin{align*}
& (\bm{h}_n - \bm{h}_{n^{\star}})^\T \bm{W}^\T \bm{W} (\bm{H}\bm{c}_\ell - \bm{h}_\ell) > \epsilon_{n^{\star}} - \epsilon_n \\
    \Longleftrightarrow & (\bm{h}_n - \bm{e}_j)^\T \bm{W}^\T \bm{W} (\bm{H}\bm{c}_\ell - \bm{h}_\ell) > \epsilon_{n^{\star}} - \epsilon_n,  \numberthis \label{eq:noise_does_not_change_order}
\end{align*}
which will leads to that the FW algorithm selects $n^\star$ at the current iteration with $\c_\ell$.

We will prove \eqref{eq:noise_does_not_change_order} using the following two Lemmas. Their proofs are provided in Appendix~\ref{app:newlemma3} and Appendix~\ref{app:newlemma4} in the supplementary material.
\begin{Lemma}\label{lemm:newlemm3}
    For any $\ell \in [N]$, $n \in [N], n\neq n^{\star}$, where ${n^{\star} := \argmax_{n} \bm{h}_n^{\T} \bm{W}^{\T}\bm{W}(\bm{H}\bm{c}_\ell - \bm{h}_\ell)}$, we have  
    \begin{equation}
(\bm{h}_n - \bm{e}_j)^{\T} \bm{W}^{\T} \bm{W} (\bm{H} \bm{c}_\ell - \mathbf{h}_\ell) \geq \dfrac{\nu \eta (1- d(\bm{H}))}{\sigma_{\max}(\bm{W})},
\label{eq:noise_does_not_change_order_step1}
    \end{equation} 
    where $\nu$ is defined in \eqref{eq:wiredcond}, and $\delta$ satisfies \eqref{eq:noisy_unreg_cond} in Theorem~\ref{theorem:noisy_unreg}.
\end{Lemma}
\begin{Lemma}\label{lemm:newlemm4}
    For any $\ell \in [N], n \in [N], n\neq n^{\star}$, where ${n^{\star} := \argmax_{n} \bm{h}_n^{\T} \bm{W}^{\T}\bm{W}(\bm{H}\bm{c}_\ell - \bm{h}_\ell)}$, we have
    \begin{equation}
        \epsilon_{n^{\star}} - \epsilon_n \leq 4(2 \gamma \delta + \delta^2).
\label{eq:noise_does_not_change_order_step2}
    \end{equation} 
\end{Lemma}

Using upper bound of noise $\delta$, we have
\begin{align*}
    \delta &\leq \sqrt{\gamma^2 
    + \dfrac{(1-d(\bm{H})) \nu \eta}{4 \sigma_{\max}(\bm{W})}} - \gamma \\
    \Leftrightarrow \delta^2 + \gamma^2 + 2 \delta \gamma 
           &\leq \gamma^2 + \dfrac{(1-d(\bm{H})) \nu \eta }{4 \sigma_{\max}(\bm{W})}  \\
    \Leftrightarrow 4(\delta^2 + 2 \delta \gamma) 
           &\leq  \dfrac{(1-d(\bm{H})) \nu \eta }{ \sigma_{\max}(\bm{W})}.
    \numberthis \label{eq:noise_does_not_change_order_step3}
\end{align*}

Using \eqref{eq:noise_does_not_change_order_step1}, \eqref{eq:noise_does_not_change_order_step2},
and \eqref{eq:noise_does_not_change_order_step3}, one can see that
\begin{align*}
    (\bm{h}_n - \bm{e}_j) ^\T \bm{W}\bm{W}(\bm{H}\bm{c}_\ell - \bm{h}_\ell) 
    &\geq \dfrac{(1-d(\bm{H})) \nu \eta }{\sigma_{\max}(\bm{W})}  \\ 
    &\geq 4(2\gamma \delta + \delta ^2) \\
    &\geq \epsilon_{n^{\star}} - \epsilon_n,
\end{align*} 
which is exactly \eqref{eq:noise_does_not_change_order}.
Until now, we have established that the FW algorithm will not update any $\C(n,:)$ such that $n\in {\cal K}^c$ before it reaches the stopping criterion. Whenever FW terminates at $\widehat{\C}$, we have
\begin{align*}
    \eta + 2 \delta 
    &\geq \norm{\bm{X} \widehat{\bm{c}}_\ell - \bm{x}_\ell}_2 \\
    &= \norm{\bm{W}(\bm{H}\widehat{\bm{c}}_\ell- \bm{h}_\ell) + \bm{V}\widehat{\bm{c}}_\ell - \bm{v}_\ell}_2 \\
    & \geq \norm{\bm{W}(\bm{H}\widehat{\bm{c}}_\ell - \bm{h}_\ell)}_2 - \norm{\bm{V} \widehat{\bm{c}}_\ell - \bm{v}_\ell}_2 \\
    & \geq \sigma_{\min}(\bm{W}) \norm{\bm{H} \widehat{\bm{c}}_\ell - \bm{h}_\ell}_2 - 2 \delta,
\end{align*}
which leads to
\[
    \norm{\bm{H}\widehat{\bm{c}}_\ell - \bm{h}_\ell}_2 \leq \dfrac{\eta + 4 \delta}{\sigma_{\min}(\bm{W})}.
\] 

This completes the proof.

\section{Proof of Theorem~\ref{theorem:identifibility}}\label{app:noisyident}
   Without loss of generality, assume that $\mathcal{K} = \set{1, \ldots , K}$ and $ \bm{H}(:, 1{:}K) = \bm{I} $. Therefore, we have
    $$\bm{C}^\star = \begin{bmatrix} \bm{H} \\ \bm{0} \end{bmatrix} $$
    that is the desired solution. Note that $\C^\star$ is an optimal solution of Problem~\eqref{eq:self}.
    Our goal is to show that Problem \eqref{eq:self-fw-fitting}'s optimal solutions are close to $\C^\star$.
    We will show this step by step.
    
\paragraph{Step 1} We first find an upper bound of the objective function associated with $\C^\star$.
    For $1\leq \ell \leq N$, it can be seen that
    \begin{align*}
        \norm{\bm{x}_\ell - \bm{X}\bm{c}_\ell^\star}_2 
        &= \norm{\bm{W}\bm{h}_\ell + \bm{v}_\ell - (\bm{W}\bm{H} + \bm{V})\bm{c}_\ell^\star}_2 \\
        & \leq \norm{\bm{W}\bm{h}_\ell - \bm{W}\bm{H}\bm{c}_\ell^\star}_2 + \norm{\bm{v}_\ell -\bm{V}\bm{c}_\ell^\star}_2 \\
        &= \norm{\bm{v}_\ell - \bm{V} \bm{c}_\ell^\star}_2 \\
        &\leq \norm{\bm{v}_\ell}_2 + \max_\ell \norm{\bm{v}_\ell}_2 \\
        & \leq 2 \max_\ell \norm{\bm{v}_\ell}_2,
    \end{align*}
    which leads to
    \[
       \norm{\bm{x}_\ell - \bm{X}\bm{c}_\ell^\star}_2^2 
        \leq 4 \max_\ell \norm{\bm{v}_\ell}_2^2 \leq 4 \dfrac{\rho}{N} \norm{\bm{V}}_{\rm F}^2.
    \] 
    Consequently, we have
    \begin{equation}
    \begin{aligned}
                         &\sum^{N}_{\ell=1} \norm{\bm{x}_\ell - \bm{X}\bm{c}_\ell^\star}_2^2 =
        \sum^{N}_{\ell=K} \norm{\bm{x}_\ell - \bm{X}\bm{c}_\ell^\star}_2^2 \\
        \leq& \sum^{N}_{\ell=K} 4 \dfrac{\rho}{N} \norm{\bm{V}}_{\rm F}^2  
        = 4\rho \dfrac{N - K}{N} \norm{\bm{V}}_{\rm F}^2 \label{eq:fit_upper_bound}      
    \end{aligned}
    \end{equation} 
    where the first equality holds because we have $\bm{x}_\ell = \bm{X}\bm{c}_\ell^\star $ for $1 \leq \ell \leq N $.
    
     By Lemma~\ref{lemma:max_smooth}, we have
    \begin{equation}
        \varPhi(\bm{C}^{\star}) \leq \norm{\bm{C}}_{\infty, 1} = K.
        \label{eq:reg_upper_bound}
    \end{equation}
    Combining \eqref{eq:fit_upper_bound} and \eqref{eq:reg_upper_bound}, we have
\begin{align*}
    \widehat{f}(\bm{C}^{\star}) 
    &= \dfrac{1}{2}\sum^{N}_{\ell=1} \norm{\bm{x}_\ell - \bm{X}\bm{c}_\ell^{\star}}_2^2 + \lambda \varPhi_{\mu}(\bm{C}^{\star}) \\
    & \leq 2 \rho \dfrac{N-K}{N} \norm{\bm{V}}_{\rm F}^2 + \lambda K. \numberthis \label{eq:obj_reg_bound}
\end{align*}
Consider any solution $\bm{C}$ such that $\widehat{f}(\bm{C}^{\star}) \geq \widehat{f}(\bm{C}) $, or $ 2\widehat{f}(\bm{C}^{\star}) \geq 2\widehat{f}(\bm{C}) $.
Then, we have
\begin{align*}
    &4 \rho \dfrac{N-K}{N} \norm{\bm{V}}_{\rm F}^2 + 2\lambda K
    \geq \norm{\bm{X} - \bm{X} \bm{C}}_{\rm F}^2 + 2\lambda \varPhi_{\mu}(\bm{C}) \\
    &\geq \norm{\bm{X}- \bm{X}\bm{C}}_{\rm F}^2 
    = \sum^{N}_{\ell=1} \norm{\bm{x}_\ell - \bm{X}\bm{c}_{\ell}}_2^2 \\
    &\geq \norm{\bm{x}_k - \bm{X} \bm{c}_k}_2^2, \quad  k \in \mathcal{K}  \\
     &= \norm{\bm{W}\bm{h}_k + \bm{v}_k - (\bm{W}\bm{H} + \bm{V})\bm{c}_k}_2^2 \quad , \quad k \in \mathcal{K} \\
    &\geq (\norm{\bm{W}\bm{h}_k - \bm{W}\bm{H}\bm{c}_k)}_2 - \norm{\bm{v}_k - \bm{V}\bm{c}_k}_2)^2 , \quad k \in \mathcal{K} \\
    &\geq (\norm{\bm{w}_k - \bm{W}\bm{H}\bm{c}_k}_2 - 2\delta)^2, \quad k \in \mathcal{K}.
\end{align*}
Therefore, we have
\begin{equation}
\sqrt{4 \rho \dfrac{N-K}{N} \norm{\bm{V}}_{\rm F}^2 + 2\lambda K} + 2\delta 
    \geq \norm{\bm{w}_k - \bm{W}\bm{H} \bm{c}_k}_2,
    \label{eq:WHc_bounded}
\end{equation} 
for any $k \in{\cal K}$.

\paragraph{Step 2} Following the idea in \cite[Lemma 17]{gillis2014robust}, define $\tau = \bm{H}(k, :) \bm{c}_k$. 
It is readily seen that $0 \leq \tau \leq 1$. 
Suppose  $\tau < 1$, one can see that for $1 \leq k \leq K$
\begin{align*}
    &\norm{\bm{w}_k - \bm{W}\bm{H}\bm{c}_k}_2 \\
    &= \norm{\bm{w}_k - [ \bm{w}_k \bm{H}(k, :) \bm{c}_k + \bm{W}(:, -k) \bm{H}(-k, :)\bm{c}_k ]}_2 \\
    &= \norm{(1- \tau)\bm{w}_k -  \bm{W}(:, -k) \bm{H}(-k, :) \bm{c}_k}_2 \\
    &= (1-\tau) \norm{\bm{w}_k -\bm{W}(:, -k) \dfrac{ \bm{H}(-k, :) \bm{c}_k}{1-\tau}}_2
\end{align*} 
where $\bm{W}(:, -k), \bm{H}(-k, :)$ refers to a sub-matrix constructed by removing the $k$th column and $k$th row from $\bm{W}$ and $\bm{H}$, respectively.

Denote $\boldsymbol \theta = \dfrac{1}{1-\tau} \bm{H}(-k, :)\bm{c}_k$, it is easy to verify that $\bm{1}^\T \boldsymbol \theta = 1, \boldsymbol \theta \geq 0$. Then,
\begin{equation}
    \norm{\bm{w}_k - \bm{W}\bm{H}\bm{c}_k}_2 \geq (1 - \tau) \kappa(\bm{W}) \label{eq:WHc_is_closed}
\end{equation} 
by the definition of $\kappa(\bm{W})$ for $k\in{\cal K}$.

Combining \eqref{eq:WHc_bounded}-\eqref{eq:WHc_is_closed}, we have
\begin{equation}
     \Rightarrow \tau \geq 1 - \dfrac{\sqrt{4 \rho \dfrac{N-K}{N}\norm{\bm{V}}_{\rm F}^2 + 2\lambda K} + 2\delta}{\kappa(\bm{W})} 
\end{equation}
In case $\tau = 1$, the inequality above holds trivially.

\paragraph{Step 3}
Since $h_{k, k} = 1$ for $k \in {\cal K}=[K]$, we have
\begin{align*}
    \tau &= \bm{H}(k, :) \bm{c}_k 
    = \sum^{N}_{\ell=1} h_{k, \ell} c_{\ell, k}
    = h_{k, k} c_{k, k} + \sum_{\ell \neq k} h_{k, \ell} c_{\ell, k} \\
         &=  c_{k, k} + \sum_{\ell \neq k} h_{k, \ell} c_{\ell, k} \numberthis \label{eq:h_at_least_1} \\
         &\leq c_{k, k} + (1-c_{k,k}) \max_{\ell \neq k} h_{ k, \ell } \leq 1.  \numberthis \label{eq:eta_bound}
\end{align*}
By \eqref{eq:eta_bound}, for $k \in{\cal K}$, we have
\[
    \tau 
    \leq c_{k , k } + (1- c_{k , k }) \max_{\ell \neq k } h_{k , \ell}
    \leq c_{k , k } + (1- c_{k , k }) d(\bm{H}),
\] 
Hence, it can be seen that for $k \in{\cal K}$, the following holds:
\begin{align*}
    \norm{\bm{C}(k, :)}_{\infty} 
    \geq \bm{c}_{k, k}
    &\geq \dfrac{\tau - d(\bm{H})}{1 - d(\bm{H})}\\
    &\geq 1 - 
\underbrace{\dfrac{\sqrt{4 \rho \dfrac{N-K}{N} \norm{\bm{V}}_{\rm F}^2 + 2\lambda K} + 2\delta}{\kappa(\bm{W})(1-d(\bm{H}))}}_{\beta}.
\end{align*} 
Meanwhile, for $1 \leq n \leq N$, we have
\begin{equation*}
    \norm{\bm{C}(n, :)}_{\infty} - \mu \log(N) \leq \varphi_{\mu}(\bm{C}(n, :))  \quad \text{( Lemma~\ref{lemma:max_smooth})}.
\end{equation*} 
This leads to
\begin{align*}
    & \sum^{N}_{n=1} \norm{\bm{C}(n, :)}_{\infty} - \mu N \log(N)
    \leq \dfrac{1}{\lambda}  \lambda \sum^{N}_{n=1} \varphi_{\mu}(\bm{C}(n, :)) \\
    &\leq \dfrac{1}{\lambda} \widehat{f}(\bm{C}^\star) \leq 
    2 \rho \dfrac{N-K}{\lambda N}\norm{\bm{V}}_{\rm F}^2 + K,
\end{align*}
where the last inequality was established in Step 1.

Finally, for $n \notin \mathcal{K}$
\begin{align*}
    &\norm{\bm{C}(n, :)}_{\infty} 
    \leq \sum_{i \in \mathcal{K}^{c}}  \norm{\bm{C}(i, :)}_{\infty}  \\
    &= \sum_{i \in [N]} \norm{\bm{C}(i, :)}_{\infty} - \sum_{i \in \mathcal{K}} \norm{\bm{C}(i, :)}_{\infty} \\
    &\leq \left( 2\rho \dfrac{N-K}{\lambda N} \norm{\bm{V}}_{\rm F}^2 + K + \mu N \log(N) \right) - K (1-\beta) \\
    &= 2\rho \dfrac{N-K}{\lambda N} \norm{\bm{V}}_{\rm F}^2 + \mu N \log(N) + \beta K.
\end{align*} 

This completes the proof.

\section{Proof of Theorem~\ref{theorem:reg_mem}}
\label{sub:proof_reg_mem}
By theorem's assumption, ${\rm supp}(\bm{c}_\ell^{\rm init}) \subseteq \mathcal{K}$ holds at initialization. Our goal is to prove that if ${\rm supp}(\bm{c}_\ell^t) \subseteq \mathcal{K}$  holds, then ${\rm supp}(\bm{c}_\ell^{t+1}) \subseteq \mathcal{K}$ always holds.

To proceed, we will need the following lemmas:
\begin{Lemma}\label{remark:arbitrarily_small}
    Denote 
    $y_{n, \ell} = \nicefrac{\exp{c_{n,\ell} / \mu}}{ \sum^{N}_{i=1} \exp{c_{n, i} / \mu}}$ where $\mu>0$.
    If $c_{n, \ell}$ is not the largest element in row $\bm{C}(n, :)$, i.e., 
    \begin{equation}
        \label{eq:not_largest_cond}
    c_{n, \ell} < \max_{i} c_{n, i},
    \end{equation}
    then we have
$
    y_{n, \ell} < (\nicefrac{1}{\abs{\mathcal{L}_{n}}}) \exp{(c_{n, \ell} - c_{n,\star})/ \mu} 
$
where $c_{n, \star} \coloneqq \max_{i} c_{n, i}$, and $ \mathcal{L}_n \coloneqq \set{i | c_{n,i} = c_{n, \star}}$.
\end{Lemma}

\begin{Lemma}
    \label{lemma:dH_relation}
    Given vector $\bm{x} \in \mathbb{R}^{K}$. Suppose  $\bm{x}$ satisfies the followings:
    $\bm{x} \geq 0, \bm{1}^\T \bm{x} = 1, \norm{\bm{x}}_{\infty} \leq a$ for some $a \leq 1$, then
\[
    \norm{\bm{x}}_2 \leq \sqrt{\max \left( 2(a-\nicefrac{1}{2})^2+\nicefrac{1}{2}, 2(\nicefrac{1}{2} - \nicefrac{1}{K})^2 + \nicefrac{1}{2} \right)}
\] 
\end{Lemma}

\begin{Lemma}
    \label{lemma:bluebluebound}
    For $n \in \mathcal{K}, m \in \mathcal{K}^{c}$, the following holds:
    \begin{equation}
    \label{eq:blueblue}
    \begin{aligned}
        (\bm{h}_m &- \bm{h}_n)^\T  \bm{W}^\T \bm{W} (\bm{H}\bm{c}_\ell - \bm{h}_\ell) \\
        &\geq - (d'(\bm{H})^2 + 2d'(\bm{H})+5)\lambda_{\max}(\bm{W}^\T \bm{W})/2.
    \end{aligned} 
    \end{equation}
    where 
    \[
    d'(\bm{H}) = \sqrt{\max(2(d(\bm{H}) - \nicefrac{1}{2})^2 +\nicefrac{1}{2}, 2(\nicefrac{1}{2} - \nicefrac{1}{K})^2+ \nicefrac{1}{2})}.
\]
\end{Lemma}
Proofs of Lemmas \ref{remark:arbitrarily_small}, \ref{lemma:dH_relation}, and \ref{lemma:bluebluebound} are relegated to the supplementary material in Appendices \ref{app:proof_arbitrarily_small}, \ref{app:proof_dH_relation}, and \ref{app:proof_bluebluebound}, respectively.

Let us assume $\C^t=\C$. The gradient of \eqref{eq:self-fw-fitting} is $$ \nabla \widehat{f}(\C) = {\bm{X}^\T (\bm{X}\bm{C} - \bm{X})} + \lambda {\nabla \varPhi_{\mu} (\bm{C})} .$$ 
Denote $\bm{P}=\bm{X}^\T (\bm{X}\bm{C} - \bm{X})$ and $\bm{Y} = \nabla \varPhi_{\mu}(\bm{C}).$ One can see that
\begin{align*}
    \bm Y(n,\ell)=y_{n, \ell} = \dfrac{\exp{c_{n,\ell} / \mu}}{ \sum^{N}_{i=1} \exp{c_{n, i} / \mu}}.
\end{align*}
Our goal is to show that there always exists $n \in \mathcal{K}$ such that $c_{n, \ell}$ satisfies condition \eqref{eq:not_largest_cond} in Lemma~\ref{remark:arbitrarily_small} for any $\ell$. If this holds, then the corresponding gradient value $ [\nabla \widehat{f}(\C)]_{n,\ell}$ is expected to be small, since the corresponding $y_{n,\ell}$ is small.

To this end, denote $\bm g_\ell$ as the $\ell$th column of $\nabla \widehat{f}(\C)$, i.e., $$\bm{g}_\ell = \bm{p}_\ell + \lambda \bm{y}_\ell,$$
where $\bm p_\ell$ and $\bm y_\ell$ are the $\ell$th columns of $\bm P$ and $\bm Y$, respectively.
Our objective then amounts to showing that $j\in {\cal K}$ where
\[
j=\argmin_{n\in [N]} \; p_{n,\ell} + \lambda y_{n,\ell}.
\]
Again, w.o.l.g., we assume that $\mathcal{K} =[K]$ and $\bm h_n=\bm e_n$ for $n\in {\cal K}$.
We use a contradiction to show our conclusion.
Suppose that for every $n \in \mathcal{K}$, $c_{n, \ell}$ is the largest element in row ${\bm{C}(n, :)}$---i.e., $c_{n,\ell}\geq c_{n,\ell'}$ for all $ \ell' \neq \ell$ for every $n\in{\cal K}$.
Since $c_{n, \ell}$ is the largest element,  and since $\C(n,:)$ is not a constant by our assumption,
one can always find an $\ell'$ such that  $c_{n, \ell} > c_{n, \ell'}$. Then, we have 
\begin{align*}
    &1 = \mathbf{1}^\T \bm{c}_\ell 
    = \sum^{N}_{i=1} c_{i,\ell} 
    = \sum^{K}_{i=1} c_{i,\ell}\\
    &= c_{n,\ell} + \sum^{K}_{i \neq n} c_{i,\ell}
    \geq c_{n,\ell} + \sum^{K}_{i \neq n} c_{i, \ell'}
    > \sum^{K}_{i=1} c_{i, \ell'}  = 1.
\end{align*} 
The third equality holds because of the assumption that ${\rm supp}(\bm{c}_{\ell}) \subseteq \mathcal{K}$.
The above is a contradiction, which means that for any given $\ell$, there must be at least an $n\in {\cal K}$ such that $c_{n,\ell} < c_{n,\ell'}$ for a certain $\ell'$. Therefore, by Lemma~\ref{remark:arbitrarily_small}, we have the following inequality:
\begin{equation}
    y_{n, \ell} < \dfrac{1}{\abs{\mathcal{L}_{n}}} \exp{(c_{n, \ell} - c_{n,\star})/ \mu} 
    < \exp{-\psi/ \mu}.
    \label{eq:y_is_small}
\end{equation} 
In the meantime, for $m \in \mathcal{K}^{c}$, we have
\[
     y_{m, \ell} = \dfrac{\exp{c_{m, \ell} / \mu }}{ \sum^{N}_{i=1} \exp{c_{m, i}/ \mu}} = \dfrac{1}{N},
\] 
because $\bm C(m,:)=\bm 0^\T$.
For an $n \in \mathcal{K}$ that satisfies \eqref{eq:y_is_small} and an $m \in \mathcal{K}^{c}$, 
we have
\begin{align*}
g_{m, \ell} - g_{n, \ell} 
&= p_{m, \ell} + \lambda y_{m, \ell} - p_{n, \ell} - \lambda y_{n, \ell} \\
&= p_{m, \ell} - p_{n, \ell} + \lambda \left(\dfrac{1}{N} - y_{n, \ell} \right).
\end{align*} 
Using Lemma~\ref{lemma:bluebluebound}, we can establish an lower bound of $p_{m, \ell} - p_{n, \ell}$,
i.e.,
\begin{align*}
    &p_{m, \ell} - p_{n, \ell} 
    = (\bm{h}_m - \bm{h}_n)^\T \bm{W}^\T \bm{W} (\bm{H} \bm{c}_\ell - \bm{h}_\ell) 
    + (\epsilon_m - \epsilon_n) \\
    &\geq (\bm{h}_m - \bm{h}_n)^\T \bm{W}^\T \bm{W} (\bm{H} \bm{c}_\ell - \bm{h}_\ell) 
    - 4(2 \gamma \delta + \delta^2 )  \\
    &\geq -(d'(\bm{H})^2 + 2d'(\bm{H})+5) \lambda_{\max}(\bm{W}^\T \bm{W})/2 
    - 4(2 \gamma \delta + \delta^2 ) 
\end{align*} 
where the first equality is by \eqref{eq:gradient_noise}, the first and second inequalities are by \eqref{eq:noise_does_not_change_order_step2} and Lemma~\ref{lemma:bluebluebound}, respectively.

Therefore, we have
\begin{equation}
    \label{eq:difference_in_gradient}
\begin{aligned}
    g_{m, \ell} - g_{n, \ell} 
    &\geq \lambda (1/N - y_{n, \ell}) \\
    &   \quad \quad -(d'(H)^2 + 2d'(\bm{H})+5) \lambda_{\max}(\bm{W}^\T \bm{W})/2 \\
    & \quad \quad -4(2 \gamma \delta + \delta^2 )  \\
    &> \lambda/N - \lambda \exp{-\psi/\mu} \\
    & \quad \quad - (d'(H)^2 + 2d'(\bm{H})+5) \lambda_{\max}(\bm{W}^\T \bm{W})/2 \\
    &  \quad  \quad  -  4(2 \gamma \delta + \delta^2 )  \\
    &\geq 0. 
\end{aligned}
\end{equation}
where the last inequality can be derived from noise bound given in \eqref{eq:noise_bound_reg}.

Hence, one can see that
\[ \argmin_n \; p_{n,\ell} + \lambda y_{n,\ell} \in \mathcal{K}\]
As a result, the update rule of FW will make ${\rm supp}(\bm{c}_\ell) \subseteq \mathcal{K}$ for the next iteration. This completes the proof.

\clearpage

{\bf Supplementary Materials of ``Memory-Efficient Convex Optimization for Self-Dictionary Separable Nonnegative Matrix Factorization: A Frank-Wolfe Approach
''} 
\begin{center}
Tri Nguyen, Xiao Fu, and Ruiyuan Wu    
\end{center}

\section{Proof of Lemma~\ref{lemm:newlemm3}}\label{app:newlemma3}

By the stopping criterion, before FW terminates, $\eta + 2 \delta \leq \norm{\bm{X}\bm{c}_\ell - \bm{x}_\ell}_2$ holds. Hence, the following chain of inequalities holds:
        \begin{align*}
            \eta + 2\delta  & \leq \norm{(\bm{W}\bm{H} + \bm{V} )\bm{c}_\ell - \bm{W}\bm{h}_\ell - \bm{v}_\ell}_2  \\
            &= \norm{\bm{W}(\bm{H}\bm{c}_\ell - \bm{h}_\ell) + \bm{V}\bm{c}_\ell - \bm{v}_\ell}_2  \\
            &\leq \norm{\bm{W}(\bm{H}\bm{c}_\ell - \bm{h}_\ell)}_2 + \norm{\bm{V}\bm{c}_\ell - \bm{v}_\ell}_2  \\
            &\leq \norm{\bm{W}(\bm{H}\bm{c}_\ell - \bm{h}_\ell)}_2 + \norm{\bm{V}\bm{c}_\ell}_2 + \norm{\bm{v}_\ell}_2 \\
            &\leq \norm{\bm{W}(\bm{H}\bm{c}_\ell - \bm{h}_\ell)}_2 + \norm{\bm{V}\bm{c}_\ell}_2 + \delta \\
            &\leq \norm{\bm{W}(\bm{H}\bm{c}_\ell - \bm{h}_\ell)}_2 + 2\delta  \\
            &\leq \sigma_{\max}(\bm{W}) \norm{\bm{H}\bm{c}_\ell - \bm{h}_\ell}_2 + 2\delta,
        \end{align*}
        \begin{equation}
            \label{eq:obj_upper_bound}
            \Longrightarrow \norm{\bm{H}\bm{c}_\ell - \bm{h}_\ell}_2 
            \geq \dfrac{\eta}{\sigma_{\max}(\bm{W})}.
        \end{equation} 
        The last inequality holds since 
        $\norm{\bm{V}\bm{c}_\ell}_2 \leq \sum^{N}_{i=1} c_{i,\ell}
        \norm{\bm{v}_i}_2 \leq \max_j \norm{\bm{v}_j}_2 = \delta $.

        Let $\bm{q}_\ell = \bm{W}^\T \bm{W} (\bm{H}\bm{c}_\ell -\bm{h}_\ell)$. In addition, w.o.l.g., let $q_{j,\ell}, q_{s,\ell}$ be the smallest and the second smallest elements in $ \bm{q}_\ell$, respectively. 
By the definition of $d(\bm{H})$, we have
    \begin{align*}
        \bm{h}_n^\T \bm{q}_\ell 
    &= \sum^{K}_{k=1} h_{k, n} q_{k,\ell} \\
    &= h_{j, n} q_{j \ell} + h_{s, n} q_{s, \ell} + \sum_{k \neq j, k\neq s} h_{k, n} q_{k,\ell} \\
    &\geq h_{j, n} q_{j, \ell} + h_{s, n} q_{s, \ell} + q_{s,\ell} \sum_{k \neq j, k\neq s} h_{k, n}  \\
    &= h_{j, n} q_{j, \ell} + h_{s, n} q_{s, \ell} + q_{s,\ell} (1-h_{j, n} - h_{s,n}) \\
    &= h_{j, n} q_{j, \ell} + q_{s,\ell} (1-h_{j, n}) \\
    &\geq d(\bm{H}) q_{j, \ell} + (1-d(\bm{H})) q_{s, \ell}, \numberthis \label{eq:hntq}
    \end{align*}
where the last inequality holds because
$$
  (d(\bm{H})-h_{j, n}) q_{s, \ell} \geq (d(\bm{H})-h_{j, n}) q_{j, \ell}. 
$$

By the definition of $\nu$ in~\eqref{eq:wiredcond}, we have
\begin{align*}
    &q_{s,\ell} - q_{j,\ell} \\
    &= \norm{\bm{H}\bm{c}_\ell - \bm{h}_\ell}_2 \left(  \min_{j \neq i} \min_{i} (\bm{w}_j - \bm{w}_i)^\T \bm{W} 
    \dfrac{\bm{H}\bm{c}_\ell - \bm{h}_\ell}{\norm{\bm{H}\bm{c}_\ell - \bm{h}_\ell}_2} 
    \right)\\
    &\geq \dfrac{\nu \eta}{\sigma_{\max}(\bm{W})}. \numberthis \label{eq:qs_qj}
\end{align*}
Consequently, we have 
\[
    \bm{e}_{j}^\T  \bm{W}^\T \bm{W} (\bm{H}\bm{c}_\ell - \bm{h}_\ell) = \bm{e}_{j}^\T \bm{q}_\ell = q_{j,\ell}.\]
In addition, since for any $n\neq j$, we have
\begin{align*}
        &\bm{h}_{n}^\T  \bm{W}^\T \bm{W} (\bm{H}\bm{c}_\ell - \bm{h}_\ell) 
        = \bm{h}_{n}^\T\bm{q}_\ell \\
        &\geq 
        d(\bm{H}) q_{j, \ell} + (1-d(\bm{H})) q_{s, \ell},
\end{align*}
which is by \eqref{eq:hntq},
\color{black}
we get 
\begin{align*}
 (\bm{h}_n - \bm{e}_j)^\T \bm{W}^\T \bm{W}&(\bm{H} \bm{c}_\ell - \bm{h}_\ell) \\
 &\geq d(\bm{H}) q_{j, \ell} + (1-d(\bm{H})) q_{s, \ell} - q_{j,\ell} \\
 &= (1-d(\bm{H})) (q_{s,\ell} - q_{j,\ell}) \\
 & \geq  \dfrac{\nu \eta (1-d(\bm{H}))}{\sigma_{\max}(\bm{W})}
 ,
\end{align*} 
where the last step is by \eqref{eq:qs_qj}.

\section{Proof of Lemma \ref{lemm:newlemm4}}\label{app:newlemma4}

For $1\leq \ell \leq N$, we have
    \begin{align*}
        \abs{\epsilon_n} &= | \bm{h}_n^\T \bm{W}^\T (\bm{V}\bm{c}_\ell - \bm{v}_\ell)  \\
                  & \qquad \qquad +\bm{v}_n^\T (\bm{W}\bm{H}\bm{c}_\ell + \bm{V}\bm{c}_\ell -\bm{W}\bm{h}_\ell - \bm{v}_\ell)| \\
        &\leq \abs{\bm{h}_n^\T \bm{W}^\T \bm{V}\bm{c}_\ell} + \abs{\bm{h}_n^\T \bm{W}^\T \bm{v}_\ell} + \abs{\bm{v}_n^\T \bm{W}(\bm{H}\bm{c}_\ell - \bm{h}_\ell)} \\
        &\qquad \qquad + \abs{\bm{v}_n^\T \bm{V}\bm{c}_\ell} + \abs{\bm{v}_n^\T \bm{v}_\ell} 
    \end{align*}
Note that we have
\begin{align*}
    \abs{\bm{h}_n^\T \bm{W}^\T \bm{V}\bm{c}_\ell} &
    \leq \norm{\bm{W}\bm{h}_n}_2 \norm{\bm{V}\bm{c}_\ell}_2 \\
    &\leq  \left(  \max_{k} \norm{\bm{w}_k}_2 \right) \left(  \max_{i} \norm{\bm{v}_i}_2 \right) \\
    &= \gamma \delta.
\end{align*} 
and
\begin{align*}
    \abs{\bm{h}_n^\T \bm{W}^\T \bm{v}_\ell} 
    &\leq \norm{\bm{W} \bm{h}_n}_2 \norm{\bm{v}_\ell}_2 \\
    &\leq \left( \max_k \norm{\bm{w}_k}_2 \right) \left( \max_i \norm{\bm{v}_i}_2 \right) 
    = \gamma \delta.
\end{align*} 
In addition, it is seen that
\begin{align*}
    \abs{\bm{v}_n^\T \bm{W} (\bm{H}\bm{c}_\ell - \bm{h}_\ell)} 
    &\leq \norm{\bm{v}_n}_2 \norm{ \bm{W} (\bm{H}\bm{c}_\ell - \bm{h}_\ell)}_2 \\
    &\leq \norm{\bm{v}_n}_2 (\norm{\bm{W}\bm{H}\bm{c}_\ell}_2 + \norm{\bm{W}\bm{h}_\ell}_2) \\
    &\leq 2 \gamma \delta.
\end{align*} 
We also have
\[
\abs{\bm{v}_n^\T \bm{V}\bm{c}_\ell} \leq \norm{\bm{v}_n}_2 \norm{\bm{V}\bm{c}_\ell}_2 \leq \delta^2,~
\abs{\bm{v}_n^\T \bm{v}_\ell} \leq \norm{\bm{v}_n}_2 \norm{\bm{v}_\ell}_2 \leq \delta^2.
\] 
Combining the upper bounds, we have
\[
    \abs{\epsilon_n} \leq 2(2 \gamma\delta + \delta^2),
\] 
and therefore, 
\begin{equation}
    \epsilon_{n^{\star}} - \epsilon_n \leq \abs{\epsilon_{n^{\star}} - \epsilon_n} \leq 2\abs{\epsilon_n} \leq 4(2 \gamma \delta + \delta^2).
\end{equation}

\section{Proof of Lemma \ref{remark:arbitrarily_small}} \label{app:proof_arbitrarily_small}
    Let $\mathcal{L}_{n}$ be a set of indices of the largest elements in row  ${\bm{C}(n, :)}$: $\mathcal{L}_{n} \coloneqq \set{\argmax_i \; c_{n,i}}$. 

    Let $c_{n,\star}$ be the maximum value in row ${\bm{C}(n, :)}$, i.e., $c_{n,\star} = c_{n, i}$ where  $ i \in \mathcal{L}_{n}$
\begin{align*}
    y_{n, \ell}
    &= \dfrac{\exp{c_{n, \ell}/\mu}}{ \sum^{N}_{i=1} \exp{c_{n, i}/\mu}}  \\
    &= \dfrac{\exp{c_{n, \ell}/\mu}}{ \sum_{i \in \mathcal{L}_{n}} \exp{c_{n,\star}/\mu}  + \sum_{i \notin \mathcal{L}_{n}} \exp{c_{n, i}/\mu} }\\   
    &= \dfrac{1}{\abs{\mathcal{L}_{n}} \exp{c_{n,\star}/\mu - c_{n, \ell}/\mu} + \sum_{i \notin \mathcal{L}_{n}} \exp{c_{n, i}/\mu- c_{n, \ell}/\mu} } \\
    &< \dfrac{1}{\abs{\mathcal{L}_{n}} \exp{c_{n,\star}/\mu - c_{n,\ell}/\mu}} \\
    &= \dfrac{1}{\abs{\mathcal{L}_{n}} } \exp{(c_{n,\ell}- c_{n, \star})/\mu },
\end{align*}
That concludes 
\begin{equation}
 y_{n,\ell} < \dfrac{1}{\abs{\mathcal{L}_{n}} } \exp{(c_{n,\ell}- c_{n,\star})/\mu }.
\label{eq:softmax-property-1}
\end{equation}

\section{Proof of Lemma \ref{lemma:dH_relation}} \label{app:proof_dH_relation}
Suppose that $x_1 = \max_{i} x_i$, which is without loss of generality. It is easy to verify $a \geq x_1 \geq 1/K$. Therefore,
    \begin{align*}
        \norm{\bm{x}}_2^2 
        = \sum^{K}_{i=1} x_i^2 
        &=  x_1^2 + \sum^{K}_{i=2} x_i^2 \\
        &\leq x_1^2 + \left( \sum^{K}_{i=2} x_i \right)^2 \\
        &= x_1^2 + (1- x_1)^2 \\
        &= 2(x_1 - \dfrac{1}{2})^2 + \dfrac{1}{2} \\
        &\leq \max \left(  2(a-\dfrac{1}{2})^2+\dfrac{1}{2}, 2(\dfrac{1}{K} - \dfrac{1}{2})^2+\dfrac{1}{2}\right)
    \end{align*}

\section{Proof of Lemma \ref{lemma:bluebluebound}} \label{app:proof_bluebluebound}
    Let $\bm{a} = \bm{h}_{m} - \bm{h}_n, \quad \bm{b} = \bm{H}\bm{c}_\ell - \bm{h}_\ell$. The LHS of \eqref{eq:blueblue} is
    \begin{align*}
        (\bm{h}_{m} - \bm{h}_m)^\T \bm{W}^\T \bm{W} (\bm{H} \bm{c}_\ell -\bm{h}_\ell) 
        &= \bm{a}^\T \bm{W}^\T \bm{W} \bm{b}.
    \end{align*}
    Note that
    \begin{align*}
        \bm{a}^\T \bm{W}^\T \bm{W} \bm{b} 
        &= \dfrac{1}{2} \begin{bmatrix}
            \bm{a}^\T  & \bm{b}^\T 
        \end{bmatrix} 
        \begin{bmatrix}
            \bm{0} & \bm{W}^\T \bm{W} \\
            \bm{W}^\T \bm{W} & \bm{0}
        \end{bmatrix} 
        \begin{bmatrix}
        \bm{a} \\ \bm{b}
        \end{bmatrix}  \\
        &= \dfrac{1}{2} \bm{z}^\T \bm{Z} \bm{z},
    \end{align*}
    where we denote $\bm{z} = \begin{bmatrix}
       \bm{a}  \\ \bm{b} 
       \end{bmatrix}, \bm{Z} = \begin{bmatrix}
            \bm{0} & \bm{W}^\T \bm{W} \\
            \bm{W}^\T \bm{W} & \bm{0}
        \end{bmatrix}$. 

        Since characteristic polynomial of $\bm{Z}$ is
        \begin{align*}
            \det (\bm{Z} - \lambda \bm{I}) 
            &= \det (\lambda^2 \bm{I} - \bm{W}^\T \bm{W} \bm{W}^\T \bm{W})  \\
            &= \det (\lambda \bm{I} - \bm{W}^\T \bm{W}) \det(\lambda \bm{I} + \bm{W}^\T \bm{W}),
        \end{align*} 
        hence $\lambda$ is eigenvalue of $\bm{W}^\T \bm{W}$ leads to $-\lambda , \lambda$ are eigenvalues of $\bm{Z}$.
    \begin{align*}
        2|(\bm{h}_{m} &- \bm{h}_n)^\T \bm{W}^\T  \bm{W} (\bm{H} \bm{c}_\ell -\bm{h}_\ell)| \\
            &= \bm{z}^\T  \bm{Z} \bm{z} \\
            &\leq \lambda_{\max} (\bm{G}) \norm{\bm{z}}_2^2 \\
            &= \lambda_{\max} (\bm{W}^\T \bm{W}) \norm{\bm{z}}_2^2 \\
            &= \lambda_{\max} (\bm{W}^\T \bm{W}) (\norm{\bm{h}_m - \bm{h}_n}_2^2 + \norm{\bm{H}\bm{c}_\ell - \bm{h}_\ell}_2^2)  \\
            &\leq \lambda_{\max} (\bm{W}^\T \bm{W}) ( (\norm{\bm{h}_m}_2 + \norm{\bm{h}_n}_2)^2 \\
            & \quad \quad \quad \quad \quad \quad + (\norm{\bm{H}\bm{c}_\ell}_2 + \norm{\bm{h}_\ell}_2)^2) \\
            &\leq \lambda_{\max} (\bm{W}^\T \bm{W}) ((\norm{\bm{h}_m}_2 + 1)^2 + (1+1)^2) \\
            &= \lambda_{\max} (\bm{W}^\T \bm{W}) (\norm{\bm{h}_m}_2^2 + 2\norm{\bm{h}_m}_2 + 5).
    \end{align*}
    Furthermore, using Lemma~\ref{lemma:dH_relation} on $\bm{h}_m$,
    \begin{align*}
        \norm{\bm{h}_m}_2 &\leq \sqrt{\max \left( 2(d(H)-\nicefrac{1}{2})^2+\nicefrac{1}{2}, 
        2(\nicefrac{1}{2} - \nicefrac{1}{K})^2 + \nicefrac{1}{2} \right)} \\
                          &= d'(\bm{H})
    \end{align*} 
Thus,
\begin{align*}
    2|(\bm{h}_{m} &- \bm{h}_n)^\T \bm{W}^\T \bm{W} (\bm{H} \bm{c}_\ell -\bm{h}_\ell)|  \\
            &\leq (d'(\bm{H})^2 + 2 d'(\bm{H})+5) \lambda_{\max}(\bm{W}^\T \bm{W}) \\
    \Rightarrow (\bm{h}_{m} &- \bm{h}_n)^\T \bm{W}^\T \bm{W} (\bm{H} \bm{c}_\ell -\bm{h}_\ell) \\
                            &\geq -(d'(\bm{H})^2 + 2d'(\bm{H})+5) \lambda_{\max}(\bm{W}^\T \bm{W})/2.
\end{align*}

\section{Proof of Proposition~\ref{proposition:row_is_not_constant}}\label{app:propinit}
Let $\bm{C}^{t_{\textnormal{init}}}(\ell, :)=\C^t(\ell,:)$. The FW algorithm's element-wise updating rule can be expressed as
    \begin{align*}
        c^{t+1}_{n, i} = \left( 1- \dfrac{2}{t+2} \right) c^{t}_{n, i} + \dfrac{2}{t+2}I_{n, i}^{t} \\
        c^{t+1}_{n, j} = \left( 1- \dfrac{2}{t+2} \right) c^{t}_{n, j} + \dfrac{2}{t+2}J_{n, j}^{t},
    \end{align*}
   where $I^t_{n, i}, J_{n, j}^{t}$ can only be either  $0$ or  $1$.

   Let $S_{n, (i,j)}^{t} = c^{t}_{n, i} - c^{t}_{n, j},$ and $ U_{n, (i,j)}^{ t} = I_{n, i}^{t} - J_{n, j}^{t}$.

   The updating rule in terms of $S_{i,j}^{t}$ is
    \[
        S_{n, (i,j)}^{t+1} = \dfrac{t}{t+2} S_{n, (i,j)}^{t} + \dfrac{2}{t+2} U_{n, (i,j)}^{t}.
    \] 
  \paragraph{Step 1}
  We first show that $S_{n, (i,j)}^{t}$ for $t>t_{\rm init}$ has the following relation with $S_{n,(i,j)}^{t_{\rm init}}$:
     \begin{equation}
         \frac{t(t+1)}{2}S_{n, (i,j)}^{t} = \frac{t_{\rm init}(t_{\rm init}+1)}{2} S_{n, (i,j)}^{t_{\rm init}} + \sum^{t-1}_{k=t_{\rm init}} (k+1)U_{n, (i,j)}^{k}.
         \label{eq:constant_row_C_D}
    \end{equation}

    Indeed, \eqref{eq:constant_row_C_D} can be shown by induction. Since \eqref{eq:constant_row_C_D} involves the sample tuple $n,i,j$ on both sides, and for the sake of simplicity, we omit these subscript temporarily in the following induction proof.

    To see this, let us consider the first iteration first. We have
 $S^{t_{\rm init}+1}
        = \frac{t_{\rm init}}{t_{\rm init}+2} S^{t_{\rm init}} + \frac{2}{t_{\rm init}+2} U^{t_{\rm init}}
        $. This means that
        \begin{equation*}
        \begin{split}
            \frac{(t_{\rm init}+1)(t_{\rm  init}+2)}{2} S^{t_{\rm init}+1} 
        = \frac{(t_{\rm init}+1)t_{\rm init}}{2} S^{t_{\rm init}} +\\
        (t_{\rm init}+1) U^{t_{\rm init}}.
        \end{split}
        \end{equation*}

To proceed, suppose that (\ref{eq:constant_row_C_D}) holds for any $t>t_{\rm init}+1$, i.e.,
\[
         \dfrac{t(t+1)}{2}S^t = \frac{t_{\rm init}(t_{\rm init}+1)}{2} S^{t_{\rm init}} + \sum^{t-1}_{k=t_{\rm init}} (k+1)U^k.
\] 
We consider the next iteration.
One can see that
\begin{align*}
    \dfrac{(t+1)(t+2)}{2} S^{t+1} &= \dfrac{(t+1)(t+2)}{2} \left( \dfrac{t}{t+2}S^t + \dfrac{2}{t+2}U^t \right) \\
  &= \dfrac{t(t+1)}{2} S^t + (t+1) U^t \\
  &= \frac{t_{\rm init}(t_{\rm init}+1)}{2} S^{t_{\rm init}} + \sum^{t-1}_{k=t_{\rm init}} (k+1)U^k \\
  & \quad \quad \quad \quad \quad \quad \quad \quad \quad \quad + (t+1)U^t \\
  &= \frac{t_{\rm init}(t_{\rm init}+1)}{2} S^{t_{\rm init}} + \sum^{t}_{k=t_{\rm init}} (k+1)U^k.
\end{align*}

    \paragraph{Step 2} As a result of step 1, we have

    \begin{multline}
        \label{eq:s_t}
        \abs{S^{t}_{n, (i,j)}} 
        = \dfrac{2}{t(t+1)}  \Big \lvert \frac{t_{\rm init}(t_{\rm init}+1)}{2} S^{t_{\rm init}}_{n, (i,j)} + \\
        \sum^{t-1}_{k=t_{\rm init}} (k+1)U^k_{n, (i,j)} \Big \rvert.
    \end{multline}
    Observe that the second term inside absolute operator of \eqref{eq:s_t} is an integer,
    \begin{align*}
        \abs{S^{t}_{n, (i,j)}} 
        &\geq \dfrac{2}{t(t+1)} \min_{z \in \mathbb{Z}} \abs{\frac{t_{\rm init}(t_{\rm init}+1)}{2} S^{t_{\rm init}}_{n, (i,j)} + z} \\
        &= \dfrac{2}{t(t+1)} \min_{z \in \mathbb{N}} \abs{D_{i,j}^{n}(\bm{C}^{\rm init}) - z} \\
        &\geq \dfrac{2}{T(T+1)} \min_{z \in \mathbb{N}} \abs{D_{i,j}^{n}(\bm{C}^{\rm init}) - z}, \numberthis \label{eq:s_t_2}
    \end{align*}
    where we use the definition of $D_{i,j}^{n}$ and the equality holds because both min operators results in 
    \[
        \min (\lceil D_{i,j}^{n}(\bm{C}^{\rm init})\rceil - D_{i,j}^{n}(\bm{C}^{\rm init}),
        D_{i,j}^{n}(\bm{C}^{\rm init}) - \lfloor D_{i,j}^{n}(\bm{C}^{\rm init}) \rfloor
        ),
    \] 
    where $\lceil \cdot \rceil, \lfloor \cdot \rfloor$ denotes ceiling and floor operators, resp.
    Next, we establish lower bound of $\abs{S_{n, (i,j)}^{t}}$ by considering 2 possibilities regarding to $D_{i,j}^{n}(\bm{C}^{\rm init})$:
    \begin{itemize}
        \item If $D_{i,j}^{n}(\bm{C}^{\rm init}) \in \mathbb{N}$, then 
            \[
                \begin{cases}
                    \abs{S_{n, (i,j)}^{t}} = 0 \quad \text{or} \\
                    \abs{S_{n, (i,j)}^{t}} \geq \dfrac{2}{T(T+1)} 
                \end{cases} ,
            \] 
            because both terms inside the absolute operator in \eqref{eq:s_t_2} are integers
        \item If $D_{i,j}^{n}(\bm{C}^{\rm init}) \notin \mathbb{N}$, then by the definition of $\xi$,
    \begin{align*}
        \abs{S_{n, (i,j)}^{t}} & \geq \dfrac{2\xi}{T(T+1)}.
    \end{align*}
    Such pair of $i,j$ exists for some $n^{\star} \in \mathcal{K}$, e.g.,  $i=i^{\star}, j=j^{\star}$, and hence
    \[|S_{n^{\star}, (i^\star,j^\star)}^{t}| \geq \dfrac{2 \xi}{T(T+1)} > 0.\]
    This further ensures that $\bm{C}^{t}(n^{\star}, :)$ at the $t$th iteration is not a constant row.
    \end{itemize}

    Combine two cases and take the minima, we have, for $ t_{\rm init} \leq t \leq T$, 
    \begin{align*}
        \min_{\substack{n \in \mathcal{K}, \\i,j \\ c^t_{n,i} \neq c^t_{n,j}}} 
        \abs{c^t_{n,i} - c^t_{n, j}} 
        &= \min_{\substack{n \in \mathcal{K}, \\i,j \\ S_{n,(i,j)}^{t} \neq 0}} 
         \abs{S_{n, (i,j)}^{t}}  \\
        &\geq \min \left( \dfrac{2}{T(T+1)},  \dfrac{2\xi}{T(T+1)} \right) \\
        &\geq \dfrac{2\xi}{T(T+1)}.
    \end{align*} 
    This completes the proof.

\section{Proof of Lemma~\ref{lemma:max_smooth}} \label{proof:lemmax_max_smooth}
Let $x_{\max} \coloneqq \max_{i} x_i$.
    \begin{align*}
        &\varphi_{\mu}(\bm{x}) 
        = \mu \log \left( \dfrac{1}{N} \sum^{N}_{i=1} \exp{ x_i/\mu} \right) \\
        &= \mu \log \left( \dfrac{1}{N} \exp { x_{\max}/\mu } \sum^{N}_{i=1}\exp{(x_i - x_{\max})/\mu } \right) \\
        &= \mu \left( -\log N + \dfrac{x_{\max}}{\mu}  +  \log \left( \sum^{N}_{i=1}\exp  {(x_i - x_{\max})/\mu } \right)\right) \\
        &=  -\mu \log N + x_{\max}  + \mu \log \left( \sum^{N}_{i=1} \exp{ (x_i - x_{\max})/\mu } \right)\\
        &\leq \|\x\|_\infty.
    \end{align*}
    Hence, it is seen that
    \begin{gather*}
         \lim_{\mu \rightarrow 0} \varphi_{\mu}(\bm{x}) =  x_{\max} =\|\x\|_{\infty}.
    \end{gather*} 
    In addition,
    \begin{align*}
        \varphi_{\mu}(\bm{x}) 
        &= \mu \log \left( \dfrac{1}{N} \sum^{N}_{i=1} \exp{ x_i/\mu } \right)\\
        &\geq \mu \log \left( \dfrac{1}{N} \exp{ x_{\max}/\mu } \right)
        = -\mu \log N + x_{\max}.
    \end{align*}
    This completes the proof.

\end{document}